\documentclass[aps,pre,twocolumn,showpacs,amssymb]{revtex4}
\usepackage[dvips]{graphicx}
\usepackage{dcolumn}% Align table columns on decimal point
\usepackage{bm}% bold math
\def\gs{\gtrsim}
\def\ls{\lesssim}
\def\be{\begin{equation}}
\def\en{\end{equation}}    
\def\gs{\gtrsim}
\def\ls{\lesssim}

\newcommand{\bi}[1]{\mbox{\boldmath$#1$}}

\def\p{\partial}
\def\bea{\begin{eqnarray}}
\def\ena{\end{eqnarray}}

\renewcommand{\theequation}{\arabic{section}.\arabic{equation}}

\begin{document}
\draft
\bibliographystyle{prsty}
\title{Attractive interaction and 
bridging transition between neutral colloidal  particles\\ 
due to preferential adsorption 
in  a near-critical binary mixture  }
\author{Ryuichi Okamoto$^1$  and Akira Onuki$^2$}
\address{
$^1$Fukui Institute for Fundamental Chemistry, 
Kyoto University, Kyoto 606-8103, Japan\\
$^2$Department of Physics, Kyoto University, Kyoto 606-8502, Japan}

\date{\today}

\begin{abstract} 
We examine   the solvent-mediated  interaction between two 
neutral colloidal particles due to 
 preferential adsorption in a near-critical binary mixture. 
 We take into account the renormalization effect due to 
 the critical fluctuations using the recent local functional theory 
$[$J. Chem. Phys. {\bf 136}, 114704 (2012)$]$. 
 We calculate the  free energy and the force between 
two colloidal particles as functions of 
the  temperature $T$,  the  composition far from the 
colloidal particles $c_\infty$, and the colloid separation 
$\ell$.  The  interaction 
is much enhanced  when the component favored 
by the colloid surfaces is   poor in the reservoir.
%For   $T$ close to  the bulk critical temperature $T_c$ 
For   such off-critical compositions, 
we find  a  surface of a first-order bridging transition $\ell= 
\ell_{\rm cx}(T,c_\infty)$ 
 in the $T$-$c_\infty$-$\ell$ space 
in a universal, scaled form, across  which 
 a discontinuous change occurs    between 
separated and bridged states. 
This surface starts from the bulk coexistence 
surface (CX) and  ends at  a bridging critical line $\ell= 
\ell_{ c}(T)$.  On approaching  the critical line, 
the discontinuity vanishes and the derivatives of the force 
with respect to $T$ and $\ell$ both diverge. 
Furthermore,  bridged states  continuously   
change into separated states 
if $c_\infty$  (or $T$) is varied from  a  value 
on CX  to value far from CX 
with   $\ell$ kept  smaller  than  
$\ell_c(T)$. 
\end{abstract}

%64.70.pv   Colloids
%68.35.Rh   Phase transitions and critical phenomena
%05.70.Jk   Critical Phenomena
%64.75.Xc   Phase separation and segregation in colloidal systems

\maketitle

\pagestyle{empty}

%%%%%%  introduction%%%%%%%%%%
\section{Introduction}
 
Much attention has  been paid to  the physics of fluids in restricted 
geometries \cite{Evansreview,Gelb}. 
The microscopic interactions between the  fluid molecules 
and the solid surface can greatly influence the phase transition 
behavior of the confined fluid \cite{Is}.
The  liquid phase is usually favored by the
walls in  fluids undergoing  gas-liquid phase separation, 
 while one component is preferentially attracted to the walls  
 in binary mixtures. 
In the film  geometry, narrow regions may be  
filled with the phase favored by  the  confining 
walls or may hold some fraction of 
the disfavored  phase.  Between these two states, 
there can be   a first-order phase transition, 
called   capillary condensation  \cite{Evansreview,Gelb,Butt}, 
depending  on   the  temperature $T$, 
and the reservoir chemical potential 
 $\mu_\infty$ for each given wall separation $D$.    
This phenomenon  occurs both in one-component fluids 
and  binary mixtures.

%From microscopic calculations, 
As another aspect, adsorption-induced density or 
composition disturbances are known to produce  
  an attractive interaction   
 between solid objects  
  \cite{Hansen,Evans-Hop}.
In binary mixtures, it is  amplified when the solvent 
far from these objects   is poor in the 
component  favored by the surfaces 
%and is close to the bulk two-phase coexistence 
 \cite{Evans-Hop}. 
Such solvent-mediate interactions 
 should   play an important 
 role in   
reversible aggregation of colloidal particles  in 
 near-critical  binary  mixtures at off-critical 
 compositions \cite{Beysens,Maher,Bonn,Guo}. In such situations, strong 
preferenial adsorption was observed 
by light scattering \cite{Beysens}. 
It is  worth noting that the  
colloid-wall interaction in a  near-critical  fluid has been  
measured directly  \cite{Nature2008,Nellen}.
 We mention  some theoretical papers, which  
treated the solvent-mediated colloid interaction 
in  an early stage \cite{Slu,Two,Lowen,Netz,Kaler}.

However, other interactions come 
into play in real systems. 
First, we should  account for the 
van der Waals (dispersion) interaction, which 
sometimes gives rise  to intriguing 
effects in wetting behavior 
\cite{Is,Butt,Bonnreview,Russel}. 
In this paper, we  examine  importance of 
  the van der Waals interaction as compared to  
the   adsorption-induced interaction. 
Second, in  aqueous  fluids,  
the colloid surface can be ionized 
and the counterions  and added 
ions form an electric double layer, resulting in 
 the screened Coulomb 
interaction  \cite{Is,Butt,Russel}. 
This repulsive interaction can be very strong 
 close to the surface, but it decays exponentially 
with the Debye screening length $\kappa^{-1}$. 
Third, in near-critical fluids, 
 the ion  distributions  and 
the critical fluctuations  become  
highly heterogeneous around the colloid 
surfaces \cite{Okamoto}. 
As a result,   the wetting layer formation 
 and the surface ionization are strongly coupled, 
which much complicates the colloid 
interaction.

On approaching the solvent criticality, 
the adsorption-induced interaction   becomes  long-ranged and 
  universal \cite{Okamoto,Fisher,Fisher-Yang,Gamb,Upton}, 
where the wall-induced heterogeneities extend 
 over mesoscopic length scales.
In the film geometry, some universal scaling relations are 
well-known and considerable 
efforts have been made to calculate \cite{Gamb,Upton} or 
measure \cite{Law,Nature2008,Gamb} 
the so-called Casimir amplitudes
(coefficients in universal relations)  \cite{Casimir}, 
In   these papers, near-critical fluids 
at the critical composition have mostly been treated 
 along the critical path $\mu_\infty=0$. 
On the other hand,  Macio\l ek {\it et al} 
\cite{Evans-Anna} found strong enhancement of one of the amplitudes 
in   two-dimensional 
Ising films under  applied magnetic field.  
 In accord with their finding, 
we  have recently  found  growing  
of the amplitudes  at off-critical compositions 
  \cite{OkamotoCasimir}, which  is particularly marked  
near a  first-order capillary condensation line 
in the $T$-$\mu_\infty$ plane. We have also examined 
 phase separation dynamics  around the capillary condensation 
 line \cite{Yabunaka}. 

In this paper, we  aim  to investigate 
 the interaction between two 
neutral colloidal particles due to 
 preferential adsorption 
in a near-critical binary mixture. 
We shall see that the solvent-mediated interaction 
is much enhanced  when the component favored 
by the colloid surfaces is   poor in the reservoir, 
as in the case of the Casimir amplitudes.
We also aim to examine the bridging transition between 
two colloidal particles \cite{Butt,Butt1}, which is analogous 
to the capillary condensation transition 
in a film. That is, two large particles 
(or one large particle and a plate) 
are connected by the phase  favored by the walls 
in bridged states, while they 
are disconnected  by intrusion of the disfavored phase 
in separated states. Bridged states appear 
near  the bulk coexistence curve 
as the separation distance is decreased. 
As previous papers on bridging, we mention 
numerical calculations 
of   phenomenological 
models \cite{Yeomans,Vino},  density functional theories  
\cite{Bauer,Evans-Hop},  
and a Monte Carlo study \cite{Higashi}.  
We also note that a bubble bridging  can occur  
between hydrophobic surfaces in water \cite{bubble}, 
which  is related to 
predrying of hydrophobic surfaces  \cite{Teshi}. 
Similarly, in the isotropic phase 
of liquid crystals, 
a nematic domain can appear 
between closely separated solid objects 
  \cite{Zu,Fukuda}.

The organization of this paper is as follows. 
In Sec.II,  we will summarize the results 
of the   local functional theory 
 of   near-critical binary mixtures. 
In Sec.III, we will present a theory 
on the  adsorption-induced interaction 
among colloidal particles together with  
 some simulation results. 
In Sec.IV, we will numerically 
investigate the bridging 
transition near the bulk criticality.

\section{Renormalized Ginzburg-Landau  free energy}

We consider near-critical binary mixtures 
using our  local functional  theory 
%to study the  adsorption-induced interaction, 
taking  into account the renormalization effect 
near  the bulk criticality, which is   similar  to 
the linear parametric model by Schofield {\it et al.}
\cite{Sc69,Onukibook} 
and the local functional model 
by  Fisher  {\it et al.} 
\cite{Upton,Fisher-Yang}. 
These authors treated near-critical fluids outside CX, 
while  we define our model    within CX. 
Furthermore, our model  satisfies  the two-scale-factor 
universality\cite{Onukibook}.  The critical amplitude ratios 
from our model are in fair agreement with 
reliable estimates  for Ising systems.

We assume   an   
upper  critical solution temperature   $T_c$ 
at a given   average pressure. The order parameter $\psi$ 
is proportional to $c-c_c$, 
where $c$ is the composition and $c_c$ 
is its critical value. The physical quantities 
exhibit singular dependence on $\psi$ 
and the reduced temperature,  
\be 
\tau= (T-T_c)/T_c.
\en 
Hereafter, 
$\alpha=0.110$,$\beta=0.325$,  $\gamma=1.240$, 
$\nu=0.630$,  $ \eta=0.0317$, and $ \delta=4.815$ 
are the usual critical exponents for 
Ising-like systems \cite{Onukibook}. 
At the critical composition with  $\tau>0$, 
the correlation length 
is written as  $\xi = \xi_0 \tau^{-\nu}$, 
where $\xi_0$ is a microscopic length.  
 The coexistence curve in the 
region  $\tau<0$ 
is  denoted by  CX. 
We write $\psi$   in the coexisting two phases as $\pm \psi_{\rm cx}$ with 
\be 
\psi_{\rm cx} = b_{\rm cx}|\tau|^\beta,
\en 
where  $ b_{\rm cx}$ is a constant.

We set up  the singular bulk free energy $F_b $, 
where the critical fluctuations with wave numbers 
larger than  the inverse correlation length 
 $\xi^{-1}$ have been coarse-grained or renormalized. 
Including the square gradient term, $F_b$ 
  is of the local functional form 
\cite{Fisher-Yang,Upton,OkamotoCasimir},   
\be
F_b  = \int d{\bi r}[f + \frac{1}{2}k_BT_c C|\nabla\psi|^2].  
\en 
where  the integral $\int d{\bi r}$ 
is within a  cell. 
Outside CX ($|\psi|>\psi_{\rm cx}$),  
 the singular free energy density  $f= f(\psi,\tau)$  
 is written in the  Ginzburg-Landau form,   
\be
 {f}= k_BT_c\bigg(\frac{1}{2}r\psi^2+ \frac{1}{4}u\psi^4\bigg).
\en 
We do not write  a constant term ($\propto 
|\tau|^{2-\alpha}$),which is a singular contribution for  $\psi=0$.   
In this paper, $C$  is made dimensionless. 
Then,  $\xi_0^{1/2}\psi$ is dimensionless 
and  $ b_{\rm cx}$ in Eq.(2.2) is of order $ \xi_0^{-1/2}$.
In the mean field theory, 
  $C$,   $r/\tau$, and $u$ in $F_b$  are constants 
independent of $\tau$ and $\psi$. 
In our renormalized functional theory, they depend on 
 a  nonnegative variable $w $ representing   
 the distance from the criticality in the $\tau$-$\psi$ plane. 
Outside CX, fractional powers of $w$ appear as 
\cite{Casimir-comment}
\bea 
C &=& w^{-\eta\nu},\\
r/\tau  &=&   \xi_0^{-2}w^{\gamma-1}  ,\\
u &=&  u^* \xi_0^{-1}w^{(1-2\eta)\nu}, 
\ena
where $u^*$ is a universal number 
and is set equal to $2\pi^2/9$ in our numerical 
analysis. 

From $\eta \ll 1$, we have $C \cong 1$. 
We determine  $w$  as   a function of 
$\tau$ and $\psi$ by   
\be 
w=  \tau +    (3u^* \xi_0) w^{1-2\beta}\psi^2.       
\en 
For $\psi=0$, we simply have $w=\tau$. 
For $\tau=0$, we obtain $w^\beta \propto |\psi|$,  
 leading to  the Fisher-Yang results  \cite{Fisher-Yang}: 
$\xi \propto |\psi|^{-\nu/\beta}$ and $f  \propto \xi^{-d} \propto 
|\psi|^{1+\delta}$. These authors  introduced 
the local correlation length $\xi(\psi)$ 
for  $\tau=0$.

In our scheme,  $\xi $ and the  
susceptibility $\chi$  
are related to the second derivative $f''= 
 \p^2 f/\p \psi^2 $  by 
\be 
 k_BT_cC/\xi^2=k_B T_c/\chi=f'' .
\en 
 For   $\tau>0$ and $ \psi=0$, we  find  
$\chi(\tau,0)= \xi_0^2 \tau^{-\gamma}$. On approaching 
CX ($\psi \to \psi_{\rm cx}$), 
 we require $f'= \p f/\p \psi \to 0$ to obtain  
 $b_{\rm cx}=1.50 /(3u^*\xi_0)^{1/2}$  
and $w=1.714|\tau|$. The susceptibility 
 on CX is determined by $\tau$ and is  written as    
\be 
 \chi_{\rm cx}= \chi(\tau,\psi_{\rm cx})
 = R_\chi \xi_0^2 |\tau|^{-\gamma},
\en 
with  $R_\chi = 8.82$. The correlation 
length  on CX is written as $\xi_{\rm cx}=
0.334  \xi_0|\tau|^{-\nu}$.

We also need to determine 
$f$  inside   CX 
($|\psi|<\psi_{\rm cx}$ and $\tau<0$) to discuss 
phase separation. Its simplest form is   
\be 
{f}={f_{\rm cx}}+ k_B{T_c} 
(\psi^2-\psi_{\rm cx}^2)^2/
(8{\chi_{\rm cx}}{ \psi_{\rm cx}^2 })  ,
\en 
where  $f_{\rm cx}$ is  
 the free energy density  on CX.  Then, $f$, 
$ f'$, and $ f''$ are 
continuous across CX.  
We also set  
$C=C_{\rm cx}= |\sigma_{\rm cx}\tau|^{-\eta\nu} 
$ inside CX,  which is the 
value of $C$ in Eq.(2.5) on CX.   Here, we neglect the 
thermal  fluctuations  
 longer than $\xi_{\rm cx}$.   In our applications, 
the space regions inside CX are  not wider  than 
$\xi_{\rm cx}$ and  the $\psi^4$ form  in Eq.(2.11) 
is well justified. As an example,   we may calculate 
the interface profile from  Eqs.(2.3) and (2.11), where  
 the surface tension is of the form \cite{Onukibook}, 
\be 
\sigma= 0.075 k_BT_c/\xi_{\rm cx}^2.  
\en 
We shall see another example inside CX in Fig.1.

\section{Colloidal particles in a near-critical fluid } 
\setcounter{equation}{0}

We consider identical colloidal 
particles with common  radius $a$ much larger than $ \xi_0$ 
in a near-critical binary mixture.
We seek  equilibrium profiles of $\psi({\bi r})$ 
around these large particles. We  assume 
 $\psi\to\psi_\infty$ far from them, where 
 $\psi_\infty$ is proportional to the 
composition deviation $c_\infty-c_c$ far from the colloidal particles. 
In its  calculation, we take the limit of  strong 
preferential adsorption. This   $\psi({\bi r})$  minimizes the grand potential 
$\Omega$,  giving rise to attraction 
among the colloidal particles 
\cite{Slu,Two,Lowen,Netz,Okamoto}.
Typical reduced temperatures in this paper   are from $-1$ 
to $-10$ in units of $ (\xi_0/a)^{1/\nu}$ and are very small for large $a$.  
Then the prewetting transition \cite{Bonnreview} 
may be assumed  to occur at lower temperatures. 
In fact, we realize thick adsorption layers 
in our numerical analysis.

\subsection{Equilibrium relations}

On  the cell  surface we 
assume ${\bi n}\cdot \nabla\psi=0$ 
 for simplicity,  but  on  the colloid surfaces
 we assume 
\be 
{\bi n}\cdot \nabla \psi= -h_1 /C 
\en 
where  $\bi n$ is  the  normal unit vector from the  interior 
to the exterior and  $h_1$ is a large positive  
surface field 
arising from  the short-range, fluid-surface 
interaction. In equilibrium, we minimize the grand potential, 
 consisting  of the bulk term and the surface term as 
\be 
\Omega = \int d{\bi r}\omega_{\rm loc} -k_BT_c \int dS h_1 \psi. 
\en 
Hereafter, $\int d{\bi r}$ 
is the space integral outside the colloidal particles  and 
in the cell, while $\int dS$ 
 is the surface integral on the colloid surfaces.
We define    the   grand potential density including the gradient contribution, \be 
\omega_{\rm loc}= f(\psi) -f_\infty  
-\mu_\infty(\psi-\psi_\infty) 
 + {k_BT_c} \frac{C}{2}|\nabla\psi|^2,  
\en 
where $f_\infty=f(\psi_\infty)$ and $\mu_\infty$ is 
related to $\psi_0$ by  
\be 
 \mu_\infty= f'(\psi_\infty).
\en 
In particular,  $\mu_\infty\cong 
(\psi_\infty+ \psi_{\rm cx})/\chi_{\rm cx}$   close to 
the negative branch of CX. The  
$\omega_{\rm loc}$ is nonnegative in our case,  
tending    to $0$  far from the colloidal particles. 
Minimization of  $\Omega$ yields  Eq.(3.1) as the boundary condition  and  
\be 
\frac{\delta F_b}{\delta \psi}= 
 f'(\psi)-k_BT_cC \nabla ^2\psi-{k_BT_c}\frac{C'}{2}|\nabla \psi|^2
=\mu_\infty, 
\en 
in the fluid region, where $C'(\psi)=dC/d\psi$.

In equilibrium, $\Omega$ is a function of 
the colloid centers  ${\bi R}_\alpha= 
(R_{\alpha x},R_{\alpha y}, R_{\alpha z})  
$ ($\alpha=1, 2,\cdots$). 
In Appendix A, we will derive the following equilibrium relation,  
\be
\frac{\p\Omega}{\p R_{\alpha i}}
  = \int_\alpha   dS  \sum _{j} 
(\Pi_{\psi ij}-\Pi_{\infty}\delta_{ij}) n_{\alpha j} ,
\en
where $i, j=x,y,z$.  The  integral $\int_\alpha dS$ is 
on the surface of the $\alpha$-th colloidal particle  
and ${\bi n}_\alpha= (n_{\alpha x}, n_{\alpha y}, 
n_{\alpha z})$ 
is the normal unit vector from the colloid interior 
to the exterior. The $\Pi_{\psi ij} $ is   
 the   stress tensor   
due to the order parameter deviations given by \cite{Onukibook}   
\bea 
&&\hspace{-1cm} 
\Pi_{\psi ij} = (\psi\delta F_b/\delta \psi - f- k_BT_c C |\nabla\psi|^2/2)
 \delta_{ij}\nonumber\\
&&   +k_BT_cC (\nabla_i\psi) (\nabla_j  \psi).  
\ena 
This tensor  satisfies   the relation, 
\be 
\sum_j \nabla_j \Pi_{\psi ij} = 
\psi \nabla_i (\delta F_b/\delta \psi), 
\en 
which  vanishes in equilibrium or under Eq.(3.5). 
Here, 
$\Pi_{\psi ij} \to \Pi_{\infty}\delta_{ij}$ far  from the 
colloidal particles  with 
\be 
\Pi_{\infty}= \psi_\infty \mu_\infty  - f(\psi_\infty) .  
\en 
%so  the integrnd   in Eq.(3.6)  vanishes  far from 
%the colloids.  
%Note that Eqs.(3.6)-(3.8) hold even in nonequilibrium. 
%The derivative 
%$\delta F_b/\delta \psi$ is explicitly written in Eq.(3.5).  
If  we further use Eq.(3.5), we obtain a simpler expression,  
\be  
\Pi_{\psi ij} = (\Pi_\infty 
- \omega_{\rm loc})
 \delta_{ij}   +k_BT_cC (\nabla_i\psi) (\nabla_j  \psi).  
\en

The expression (3.7) and the relation (3.8) 
are valid   even in 
nonequilibrium and have in fact been used 
in dynamics\cite{Onukibook,Yabunaka}. 
Note that the total stress tensor 
may be  expressed as $p_0\delta_{ij} 
+ \Pi_{\psi ij}$  in binary mixtures, 
where $p_0$ is a large background pressure 
 nearly uniform  in the cell 
(with small variations  arising from 
sounds and gravity).

\subsection{Scaling and strong adsorption limit} 
%%1 
\begin{figure}
\begin{center}
\includegraphics[scale=0.43]{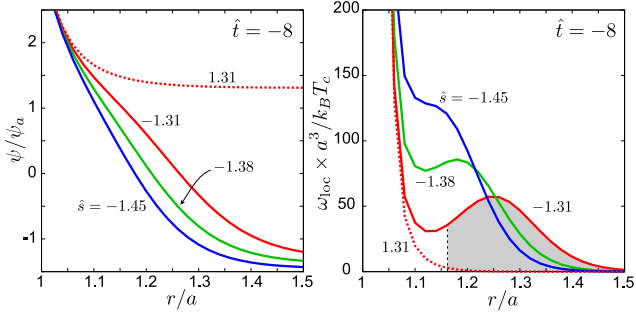}
\caption{\protect Normalized order parameter 
$\psi(r)/\psi_a$  (left) and normalized 
grand potential density $\omega_{\rm loc}(r) 
a^3/k_BT_c$ (right) vs   $r/a$ around 
a single colloid for 
 $ \hat{s}=\psi_\infty/\psi_a= 1.31, -1.31, -1.38$, 
and $-1.45$ with  $\hat{t}=\tau(\xi_0/a)^{1/\nu}= -8$. 
Shaded area below the curve of 
$ \hat{s}= -1.31$ is equal to the 
normalized surface tension $\sigma a^2/k_BT_c$. 
On CX,  $\hat{s}=\pm 1.30$ and 
$\xi=0.09 a$ for $\hat{t}=-8$.}
\end{center}
\end{figure}

We make Eq.(3.5) dimensionless by 
scaling the position $\bi r$ by $a$ and  $\psi$ 
by  $\psi_a$, where  $\psi_a$ is is  a characteristic order parameter 
around the colloidal particles of the form,  
%$(\propto a^{-\beta/\nu})$ defined by 
\be
\psi_a = (\sqrt{24}\xi_0/a)^{\beta/\nu}/(3u^* \xi_0)^{1/2}, 
\en 
Use of    $b_{\rm cx}$ in Eq.(2.2) gives     
$\psi_a= 1.47 b_{\rm cx}(\xi_0/a)^{\beta/\nu}$. 
By scaling  $\tau$ and  $\psi_\infty$, 
we introduce two  parameters, 
\bea
&& \hat{t}= \tau (a/\xi_0)^{1/\nu}, \\
&& {\hat{s}}= 
\psi_\infty /\psi_a.  
\ena 
The scaled correlation length 
 $\xi/a$  is 
given by   $\hat{t}^{-\nu}$   for  $\tau>0$ 
on the critical path, 
$0.13  |\hat{s}|^{-\nu/\beta}$ for $\tau=0$,  
 and   $0.3|{\hat{t}}|^{-\nu}$ on  CX. 
The  CX curve is expressed  as 
  $\hat{s}= \pm {\hat s}_{\rm cx}$ 
with ${\hat s}_{\rm cx}=\psi_{\rm cx}/\psi_a= 
 0.66|\hat{t}|^{\beta}$ from   Eq.(2.2).   In our calculations,  
 we may use the scaled quantities only, where 
we need not specify the ratio  $\xi_0/a \ll 1$. 
The scaling factors 
$\tau/\hat{t} =(\xi_0/a)^{1/\nu}$ 
 and  $\psi_\infty/ b_{\rm cx} \hat{s}= 1.47 (\xi_0/a)^{\beta/\nu}$. 
 are needed when our theoretical results 
are compared with experimental data. 
For example, if $a/\xi_0=10^4$, 
they are   
$0.40\times 10^{-6}$ and  
$ 0.011$, respectively.

We write    the  value of $\psi$ 
on the colloid surfaces as  $\psi_0$. 
For sufficiently large $\psi_0$, 
the near-wall behaviors  
of  $\psi$ and $\omega_{\rm loc}$ are 
expressed as    
\cite{OkamotoCasimir,Rudnick,Fisher-Yang}
\bea 
&&
\psi \sim  \xi_0^{\beta/\nu-1/2} (\lambda+\ell_0)^{-\beta/\nu}, 
\\
&& \omega_{\rm loc}  \sim k_BT_c (\lambda+\ell_0)^{-3}. 
\ena 
where $\lambda$ is the distance from 
such a surface.  We here assume  that $\lambda$ 
is  shorter than   the correlation length 
$\xi= \xi(\tau,\psi_\infty)$ far from the surface.  
The  length  $\ell_0$ is of the order 
of the local correlation length near the surface  
($\propto   \psi_0^{-\nu/\beta}$) \cite{Fisher-Yang}. 
In terms of $b_{\rm cx}$ in Eq.(2.2),  we have     
\be 
\ell_0= 0.544 \xi_0(b_{\rm cx}/\psi_0)^{\nu/\beta}.
\en   
where we assume $b_{\rm cx}^{-1} 
 \psi_0 \sim \xi_0^{1/2}\psi_0 
\ll 1$ so $\ell_0\gg \xi_0$. 
In  terms of $\psi_a$, 
we also have $\ell_0/a= 
(\beta/2\nu) (\psi_a/\psi_0)^{\nu/\beta}$.  
For $\lambda\gg \ell_0$,  $\psi$ and $\omega_{\rm loc}$ 
become independent of $\ell_0$ or $\psi_0$. 
From Eq.(3.1), we obtain 
the scaling relation,  
\be 
h_1 \sim C(\psi_0) \psi_0/\ell_0 
\sim   \psi_0^{\delta-\nu/\beta},  
\en 
where $\delta-\nu/\beta =(3-\eta)/(1+\eta)\cong 3$.
The strong adsorption  
condition  $\xi_0^{1/2}\psi_0 \gg  |\tau|^\beta$ 
is   realized with increasing   $h_1$ or on approaching 
the bulk criticality.  
 In our numerical analysis, we assume  
 $h_1 /C(\psi_0)  =170  \psi_a /a$ 
to  obtain  $\psi_0/\psi_a\sim  10$. 
See Fig.1 for the near-wall behaviors of $\psi$ 
and $\omega_{\rm loc}$ in the strong adsorption.

The  integral of 
$\psi$ in the near-wall layers  
with $0<\lambda< \ell_0$ is proportional to 
$\psi_0\ell_0 \sim \psi_0^{1-\nu/\beta}$  
and becomes  negligible  for large $\psi_0$ 
(since  $\nu/\beta \sim 2$),  
while that in the region 
$\ell_0<\lambda<\xi(\tau,\psi_\infty)$ 
grows  as 
$\xi^{1-\beta/\nu}$(critical adsorption) \cite{Bonnreview}.    
It follows a well-defined  preferential adsorption, 
\be 
\Gamma= \int d{\bi r}[\psi({\bi r})-\psi_\infty],
\en  
which  is independent of $h_1$ for large $h_1$. 
 On the other hand, 
the  integral of $\omega_{\rm loc}$ 
in the layers with $0<\lambda<\ell_0$ 
and the surface free energy in Eq.(3.2) 
($\propto h_1)$ 
are both proportional to $\psi_0^{2\nu/\beta}$ 
and are large in magnitude. However, 
they  are  constants nearly  
independent of $\tau$ and $\psi_\infty$ and 
are irrelevant in the capillary condensation 
and the bridging transition (see  discussions 
below  Eq.(3.24)), which much simplifies our results.

In the strong adsorption regime, 
the profile of $\psi$   is highly nontrivial 
for negative   $\psi_\infty $, since   
$\psi$ changes from a large positive value 
near the surface to  $\psi_\infty<0$ 
far from it.   To illustrate this aspect, we  here 
consider  the simplest case of 
a single   spherical particle   \cite{curved}, where 
$\psi(r)$ is a function of the distance 
$r$ from the particle   center.  In this case, 
if $\psi_\infty $ approaches the CX value 
$-\psi_{\rm cx} $  under the condition 
 $\xi \cong \xi_{\rm cx}\ll a$, 
 the thickness of the adsorption layer 
increases logarithmically with increasing $a$ as  \cite{curved} 
\be 
\xi_{\rm ad}= \xi\ln (a/\xi).
\en  
It is also known that  the  contribution 
to $\Omega$ from the  transition 
 region ($r-a \sim \xi_{\rm ad}$) 
is  of order   $4\pi a^2 \sigma$, 
where $\sigma$ is the surface tension in Eq.(2.12).  
%These  results   are needed to 
%explain  the bridging transition in our case. 

In Fig.1,   $\psi(r)/\psi_a$  and 
$\omega_{\rm loc}(r) a^3/k_BT_c $  are  
displayed  around a single colloidal particle  for 
 $\hat{t}=-8$, where $\hat s=\pm 1.30$ on 
 the positive and negative branches of CX..
For  ${\hat s}=1.31$,  
the adsorption layer thickness is $\xi=0.09a$. 
For  ${\hat s}=-1.31$, it 
is thicker than $\xi$ by a few times 
and is of order $\xi_{\rm ad}$ in Eq.(2.19).
Furthermore, for  ${\hat s}=-1.31$, 
 $\omega_{\rm loc}(r) a^3/k_BT_c $  
exhibits  a peak around $r-a \sim \xi_{\rm ad}$ 
with its area being about  $\sigma a^2/k_BT_c$. 
However, the peak recedes and diminishes  
for smaller $\hat s$ ($-1.38$ and $-1.45$).

%2 
\begin{figure}
\begin{center}
\includegraphics[scale=0.5]{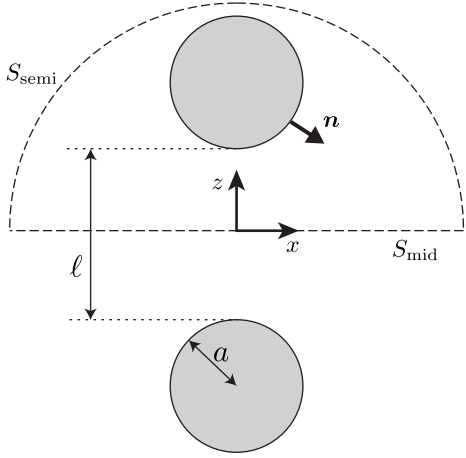}
\caption{\protect Geometry of 
 two identical spherical colloidal 
particles with radius $a$ and separation $\ell$ 
in the $xz$ plane.
Surfaces $S_{\rm mid}$ and $S_{\rm semi}$ 
are introduced in Appendix A. 
}
\end{center}
\end{figure}

%3
\begin{figure}
\begin{center}
\includegraphics[scale=0.44]{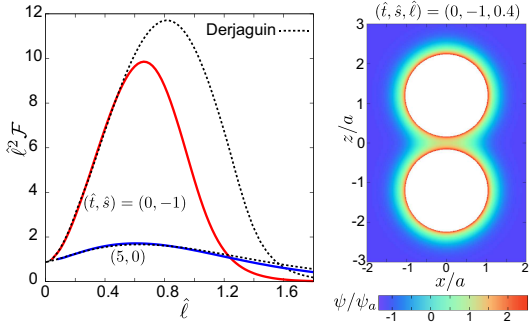}
\caption{\protect
(Color online) Left: 
${\hat\ell }^2 {\cal F}$ vs $\hat\ell=\ell/a$ 
for $(\hat{t}, \hat{s})=(0,-1)$ (red bold line) 
and $(5,0)$ (blue bold line),  
where  the correlation length $\xi$  
is  $0.13a$  for the former and $0.36a$ for the latter. 
The normalized force $\cal F$ is calculated  
from Eq.(3.24). Curves  from the Derjaguin 
approximation in Eq.(B5) 
are also written (red and blue dotted lines).
Right: Normalized order parameter $\psi(x,0,z)/\psi_a$ 
on the $xz$ plane according 
to the color bar below for 
$(\hat{t}, \hat{s},\hat{\ell})=(0,-1,0.4)$. 
For these $(\hat{t},\hat{s})$, 
 no bridging transition occurs for any $\hat\ell$. 
}
\end{center}
\end{figure}

%4
\begin{figure}
\begin{center}
\includegraphics[scale=0.44]{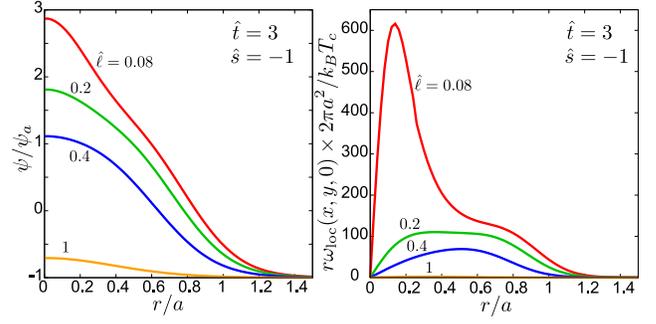}
\caption{\protect
(Color online) Midplane behaviors of 
 $\psi(r,0)/{\psi_a}$ (left) and $r 
\omega_{\rm loc}(r,0)\times  2\pi a^2/k_BT_c$ (right)  
as functions of $r/a=(x^2+y^2)^{1/2}/a$ at $z=0$   
for  $(\hat{t}, \hat{s})=(3,-1)$ with   $\xi/a= 0.12$. 
Here, ${\hat\ell}$ decreases  as $1.0$,0.4, $0.2$,  and 0.08. 
Then,     $\psi(0,0)/{\psi_a}$ at the midpoint 
is equal to  $-0.707$, 1.11, 1.81, and $2.87$, respectively, 
while ${\cal F}$ grows dramatically 
 as 1.11, 44.1, 86.5, and 219,  respectively.  
Area below each curve  (left) is equal to $\cal F$.}
\end{center}
\end{figure}

\subsection{Two colloidal particles } 

%ell=a'Ì'Æ'«A
%F_vdw / A_H =-0.005739
%ˆê•û‹ßŽ—Ž®'Ì'Ù'¤'́A-1/12=-0.083333
%W=w(a/\xi_0)^{1/\nu}, 'Æ'¨'­'ƁA'QD'VŽ®'Í
%W=\hat{t}+24^{\beta/\nu} W^{1-2\beta} \hat{s}^2
%\hat{s}=0.1'Ì'Æ'«A(W-\hat{t}) / \hat{t} <0.1'Æ'È'é'̂́A
%\hat{t} >0.34'̏ꍇ'Å'·B
%'Å'·'̂ŁA\hat{t} >0.34'È'ç'ÎW \cong \hat{t}'ÆŒ¾'Á'Ä—Ç'¢'ÆŽv'¢'Ü'·B
%  |hat{t}| > 24^{1/2\nu} |\hat{s}|^{1/\beta}  

As in Fig.2, we consider two  colloidal 
particles with equal radius $a$. In our numerical analysis, 
they are placed in the middle  of a cylindrical cell 
with radius $R_0=8a$ and height $H_0=16a$. 
The  system  is then in the region 
$0< (x^2+y^2)^{1/2}<R_0$ and $0<z<H_0$.
The particle centers are at $(0,0, \pm (\ell/2+a))$ 
with $\ell$ being the surface-to-surface 
separation distance. 
Hereafter, we set 
\be 
\hat{\ell}=\ell/a.
\en

When the system lengths ($R_0$ and $L_0$) 
much exceed $a$, it is convenient 
to write $\Omega$ as  
\be 
\Omega= \Omega_\infty -k_BT_c{\cal G},   
\en 
where $\Omega_\infty$ is the value of 
$\Omega$ for $\ell\gg a$. That is, if $\Omega_1$ 
is  the grand potential for one isolated 
colloidal particle, we have $\Omega_\infty=2\Omega_1$. 
The dimensionless quantity  $\cal G$ 
is a universal function of $\hat{t}$, 
$\hat{s}$, and $\hat{\ell}$ decaying to 0 
for large $\hat\ell$. Note that it  is independent of 
$h_1$ in the strong adsorption limit, as discussed in 
Subsec.IIIB.  
The adsorption-induced  force between the two colloidal particles  
 is given by 
\be 
\frac{\p \Omega}{\p \ell}= \frac{k_BT_c}{a}  {\cal F} .
\en 
The dimensionless  functions 
   ${\cal F}(\hat{\ell})$ and 
${\cal G}(\hat{\ell})$    are related by 
\be 
{\cal F}
= -  \frac{\p }{\p \hat{\ell}}{\cal G}.  
\en 
We also have ${\cal G}(\hat{\ell}) 
=  \int_{\hat{\ell}}^\infty d{\hat{\ell}'} 
{\cal F}(\hat{\ell}')$.  In the derivative and the integral with respect to 
$\hat\ell$,   $\hat{t}$ and $\hat{s}$ are fixed.

From the calculations in Appendix A, the normalized 
force $\cal F$ is    expressed in a convenient form,     
\be 
{\cal F}
= \frac{a}{k_BT_c} \int  dxdy~ \omega_{\rm loc}(x,y,0) .
\en
where the integral is on the $xy$ plane 
with $z=0$ (the midplane between the two colloidal particles). 
From the geometrical symmetry, $\p \psi/\p z=0$ on this plane,  
we may set $\Pi_{\psi zz}= \Pi_\infty -\omega_{\rm loc}$ 
from Eq.(3.10). Also the integral $\int dxdy$ may be replaced 
by $2\pi \int dr r$,  since $\omega_{\rm loc}(x,y,0)$ depends 
only on $r=(x^2+y^2)^{1/2}$.  If $\ell\gg \ell_0$,
the midplane is far from the transition layers 
with thickness $\ell_0$ and 
$\omega_{\rm loc}(x,y,0)$ becomes 
 independent of $\psi_0$ or  $h_1$. 
In this paper, we  thus 
calculate $\cal F$ from Eq.(3.24).

Notice that we may use Eq.(3.15)   on 
the midplane between the two colloidal particles  
for small $\ell$ \cite{mid},  
where  we set $\lambda=\ell+r^2/a\gg \ell_0$ with $r= (x^2+y^2)^{1/2}$. 
In  Eq.(3.24), the integral in the range $r \ls 
(\ell a)^{1/2}$  then becomes    
\be 
{\cal F}\sim  a\int_0^\infty 
 drr (\ell+r^2/a)^{-3} \sim 
{\hat{\ell}}^{-2}.
\en  
To be precise,  Eq.(3.24) yields 
$\lim_{\hat{\ell} \to 0}
{\hat{\ell}}^{2}{\cal F}= 0.205\pi$ as the coefficient 
in Eq.(3.25). In Appendix B, 
 the Derjaguin approximation  \cite{Is,Butt,Russel} 
for small $\hat\ell$ will yield    
\be 
{\cal G} \cong\pi \Delta_{\rm cri}
\hat{\ell} ^{-1},\quad  
{\cal F} \cong \pi \Delta_{\rm cri}
 \hat{\ell} ^{-2}, 
\en
with  $ \Delta_{\rm cri} \cong 
0.279 $. The coefficient $\pi \Delta_{\rm cri}$ 
is somewhat larger than that from Eq.(3.24).   
This  small-$\hat\ell$  behavior stems  from the de Gennes-Fisher 
theory for  near-critical films \cite{Fisher,Gamb,Nature2008}. 
Furthermore, in Appendix B,  we shall see that 
if  $\ell $ exceeds  the correlation 
length $\xi$ without bridging, 
$\cal F$ and $\cal G$ decay exponentially as  
\be
{\cal F} \sim (a/\xi)^2 
e^{-\ell/\xi}, \quad {\cal G} \sim (a/\xi)
e^{-\ell/\xi}, 
\en
 where  $\xi$  is  determined  
by $\tau$ and $\psi_\infty$ 
from  Eq.(2.9). These  relations follow 
in separated states if 
the midpoint value of $\psi$ at $z=x=y=0$ 
is close to $\psi_\infty$ \cite{OkamotoCasimir}. 

Note that Eq. (3.26) 
holds   for $\hat{\ell} \ls \xi/a$ and 
Eq.(3.27) 
for $\xi/a \ls  \hat{\ell} \ll 1$. However,  
the exponential decays in Eq.(3.27) are 
observed even for $\hat{\ell}\sim 1$ 
in  our numerical analysis (see Figs.5 and 6). 
The same exponential form  of $\cal G$ 
was found  in the previous papers 
\cite{Bonn,Gamb,Gamb1}.  
%for $\psi_\infty=0$ 
%and $\tau>0$ by Gambassi {\it et al.} 

\subsection{Numerical results without bridging transition}

In Figs.3-6, we present numerical results 
where there is no   bridging transition. 
We aim to show that $\cal G$ and  $\cal F$ are 
much more enhanced for $\hat{s}<0$ than for $\hat{s}>0$.

In the left panel of Fig.3,  
we show curves of  ${\hat\ell }^2 {\cal F}$ vs $\hat\ell$ 
calculated from Eq.(3.24) and 
  those from the   Derjaguin approximation  for 
$(\hat{\tau}, \hat{s})=(0,-1)$ and $(5,0)$. 
They  tend  to a constant as $\hat{\ell} \to 0$ 
as in Eqs.(3.25) and (3.26). Remarkably,  for ${\hat s}<0$ and ${\hat t}=0$, 
${\hat\ell }^2 {\cal F}$ increases up to of order 10 
to exhibit a peak as a function of $\hat\ell$,  where the  
 peak position is at 
$\ell 
%\sim a|\hat{s}|^{-\nu/\beta} 
\cong  6.14\xi$ from  Eq.(B10).   
On the other hand,  for  ${\hat t}>0$ and  ${\hat s}=0$, 
${\hat\ell }^2 {\cal F}$  exhibits only a 
 rounded maximum 
 of order 1 at $\ell\sim 1.64\xi$ 
from Eq.(B12). 
% These  behaviors   will be derived 
%around   Eqs.(B9)-(B11) in  
%the   Derjaguin approximation. 
 We recognize that the force is much enhanced for 
negative $\hat{s}$ and the  Derjaguin approximation nicely holds  
for $\hat{\ell} \ls 1$. 
In the right panel of Fig.3, 
 we present   $\psi(x,0,z)/\psi_a$ in gradation    for 
$(\hat{\tau}, \hat{s},\hat{\ell})=(0,-1,0.4)$ 
in the $xz$ plane, where 
$\psi$ is large  in the region  between the two 
colloidal particles.

In  Fig.4, for  $(\hat{\tau},\hat{s})=(3,-1)$, 
we show  $\psi(r,0)/\psi_a$  
and $r \omega_{\rm loc}(r,0)\times 2\pi a^2/k_BT_c$ 
vs  $r/a=(x^2+y^2)^{1/2}/a$ at $z=0$.
We change  $\hat{\ell}$  as $1$, 0.4, 0.2, and 0.008. 
For $\hat{\ell}=1$, the two collloidal particles 
are so separated  such that   
$\psi(r,0)<0$ resulting in  a small  ${\cal F}=1.1$. 
On  the other curves of smaller $\hat \ell$, 
$\psi(r,0)$ decreases from positive to negative 
with increasing $r$ 
and ${\cal F}$ increases dramatically 
up to 219. 
The behavior of the latter curves are consistent with 
the theoretical expressions:  
$
\psi \propto [\hat{\ell}+ (r/a)^2]^{-\beta/\nu}
$ and  
$ 
\omega_{\rm loc}\propto [\hat{\ell}+ (r/a)^2]^{-3},
$ at $z=0$, which follow  
 from   Eqs.(3.14) and (3.15) with  
$
\lambda =\ell+ r^2/a\gg \ell_0 
$     
as in  Eq.(3.25).

%% 5
\begin{figure}
\begin{center}
\includegraphics[scale=0.44]{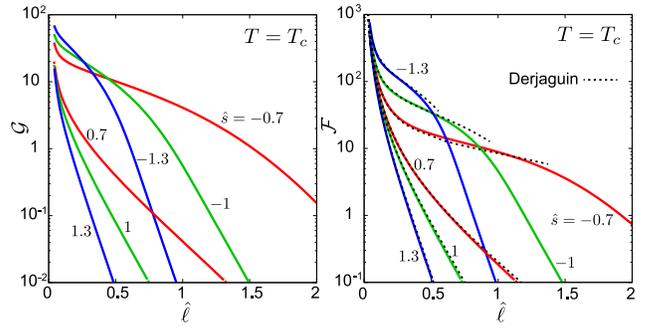}
\caption{\protect
(Color online) Normalized interaction free energy  ${\cal G}$ (left) and 
  normalized force $\cal F$ (left) vs $\hat\ell$ at $\hat{t}=0$ 
for  ${\hat s} 
=-1.3$, $-1$, $-0.7$, $0.7$, $1$ and $1.3$.    
Curves from the Derjaguin 
approximation   (dotted lines) are also written for $\cal F$ (right). 
These quantities  are much larger 
for $\hat{s}<0$ than for $\hat{s}>0$ for not very small $\hat{\ell}$.   
The slopes of the curves are close to $- a/\xi$ for relatively large $\hat\ell$  from Eq.(3.27). Here, there is no bridging transition.  }
\end{center}
\end{figure}

%% 6 
\begin{figure}
\begin{center}
\includegraphics[scale=0.44]{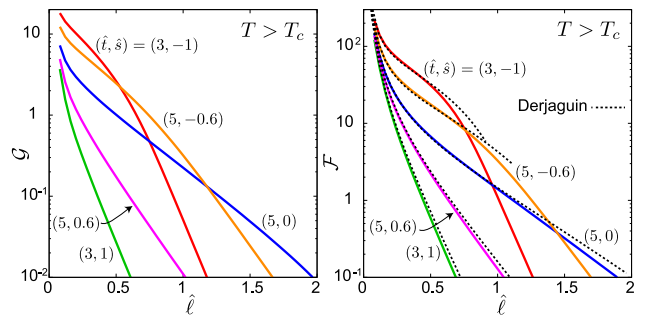}
\caption{\protect 
(Color online) 
 ${\cal G}$ (left) and 
 $\cal F$ (left) vs $\hat\ell$ 
for $(\hat{t}, \hat{s})=(3,-1)$, (5,-0.6), (5,0), (5,0.6), and (3,1) 
with     $\hat{t}>0$. Curves from the Derjaguin 
approximation  (dotted lines) are also written for $\cal F$ (right). 
As in Fig.5, they strongly depend on the sign of 
$\hat{s}$, with the slopes being  $- a/\xi$ for 
relatively large $\hat\ell$. Here, there is no bridging transition.
}\end{center}
\end{figure}

Next, we plot   ${\cal G}$  and 
  $\cal F$ vs $\hat\ell$ 
for six values of $\hat{s}$ at $\hat{\tau}=0$  in Fig.5 and 
for five values of  $(\hat{t}, \hat{s})$ 
with      $\hat{\tau}>0$ in Fig.6 on 
semi-logarithmic scales. 
 In these examples, there is no bridging transition for any $\hat\ell$.  
For small $\hat\ell$, we have the behaviors in 
Eq.(3.26).    For relatively large $\hat\ell$ 
($\xi\ll \ell \ls a$), both 
 ${\cal G}$  and   $\cal F$  decay exponentially 
as $\exp(-\ell/\xi)$. We confirm that  the slopes of these curves 
 are  close to $a/\xi$ for $\hat{\ell}\sim 1$, 
where $\xi$ is calculated from Eq.(2.9). 
We can again see  that 
  $\cal F$ is  well approximated by 
 the Derjaguin approximation for $\hat{\ell}\ls 1$.

%% 7
\begin{figure}
\begin{center}
\includegraphics[scale=0.44]{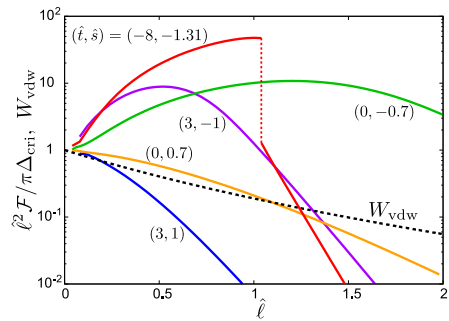}
\caption{\protect
(Color online) ${\hat\ell}^2{\cal F}/\pi\Delta_{\rm cri}$ (bold line)  
and $W_{\rm vdw}$ (dotted line) in Eq.(3.32) as functions of $\hat\ell$ 
on a semi-logarithmic scale  
for $(\hat{t}, \hat{s})=(3,-1)$, 
(0,-0.7), (0,0.7), (3,1), and (-8, -1.31).
These curves start from 1 at $\hat{\ell}=0$. 
For ${\hat\ell}\sim 1$,  ${\hat\ell}^2{\cal F}/\pi\Delta_{\rm cri}$ 
is  of order 10 for negative $\hat s$ 
without bridging  
and is even  of order 100 
at the bridging  transition, much exceeding $W_{\rm vdw}$. 
A bridging transition occurs for 
 $(\hat{t},\hat{s})=(-8, -1.31)$ 
at $\hat{\ell}=1.04$. 
}
\end{center}
\end{figure}

\subsection{Van der Waals interaction}

So far,  we have not  explicitly 
accounted  for the  pairwise van der Waals  
interaction \cite{Is}
 among constituent molecules, which was treated as one of 
the main elements causing colloid aggregation \cite{Beysens,Petit}. 
The resultant potential  $U_{\rm vdw}(r)$  between 
two colloidal particles 
 with equal radius $a$  is written as \cite{Butt,Russel}  
\be 
U_{\rm vdw}
= - \frac{A_{\rm H}}{6} \bigg [ \frac{2a^2}{r^2-4a^2} +\frac{2a^2}{r^2} 
+ \ln \bigg(1- \frac{4a^2}{r^2}\bigg)\bigg] .
\en 
where $r=2a+\ell$ is the  center-to-center distance. 
The Hamaker constant   $A_{\rm H}$ is in many cases  
of order  $10^{-19}$J, but it can 
  change its sign \cite{Is,Bonnreview} 
and can be   very small 
for  some systems of colloids and binary mixtures \cite{Bonn}.  
Without charges, the total potential is of the form, 
\be 
U_{\rm tot}= -k_BT_c {\cal G} + U_{\rm vdw}, 
\en 
consisting of  the adsorption-induced 
part and the van der Waals part. 
The former is very sensitive to $\tau$ and $\psi_\infty$ 
 in the critical ranges, 
while the latter is insensitive to them.  
If we further include the charge effects, 
we should add an appropriate 
chrage-induced interaction 
$U_C$ in Eq.(3.29) \cite{Nature2008,Okamoto,Bonn,Beysens,Gamb}
 (see item (3) in Sec.V for more discussions).

The force from the  van der Waals interaction reads   
\be 
F_{\rm vdw}    =   \frac{d}{d r}U_{\rm vdw}    
 = \frac{32A_Ha^6}{3\ell^2r ^{3}  (r+2a) ^{2}}
\en 
As  $\hat{\ell}\to 0$,  we find 
$F_{\rm vdw}  
\cong {A_{\rm H}}/{12a}{\hat{\ell}^2}$ 
This  behavior is  the same as 
that  of  $\cal F$ in Eq.(3.26).  
So we compare  the coefficients 
in front of the power ${\hat{\ell}}^{-2}$ 
of the two forces, 
 ${A_{\rm H}}/{12a}$  and 
$\pi \Delta_{\rm cri}\times k_BT_c/a$, 
 to obtain the  ratio,   
\be 
R_{\rm vdw}= {A_H}/({12\pi k_BT_c\Delta_c}), 
\en 
where the denominator is  $0.4\times 10^{-19}$J 
for $T_c  \cong 300$K. If $|A_{\rm H}|$ is smaller 
than $0.4\times 10^{-19}$J, 
we have  $|R_{\rm vdw}| < 1$ and   the van der Waals  
interaction is weaker than  the 
adsorption-induced interaction at least 
for small   ${\hat{\ell}}$.

However,   $\hat{\ell}^2{\cal F}$ 
grows for $\hat{s}<0$ with increasing 
$\hat{\ell}$ as in Fig.3,     
so we need to examine the relative importance 
of the van der Waals interaction 
and  the adsorption-induced interaction 
for larger $\hat{\ell}$. 
To this end, in Fig.7, 
we plot ${\hat\ell}^2{\cal F}/\pi\Delta_{\rm cri}$ 
for four typical cases together with 
\be 
W_{\rm vdw} \equiv  \frac{12 }{A_H a} {\ell^2} 
F_{\rm vdw}    
 = \frac{128}{(\hat{\ell} +2) ^{3}  (\hat{\ell} +4) ^{2}}
\en 
In Fig.7, while all the curves start from  
unity for   $\hat{\ell}\to  0$, 
 the normalized quantity  ${\hat\ell}^2{\cal F}/\pi\Delta_{\rm cri}$  
increases up to a maximum about 
10 for $\hat{s}<0$ without 
bridging formation  and 
can even be of order 100 close to a 
bridging transition  with increasing  $\hat{\ell}$. 
Thus, at an  off-critical composition with 
 $\hat{s}<0$,  the adsorption-induced 
interaction can  well dominate 
over the van der Waals interaction 
(even for  $|A_{\rm H}|\sim  10^{-19}$J).

%% 8
\begin{figure}[t]
\begin{center}
\includegraphics[scale=0.75]{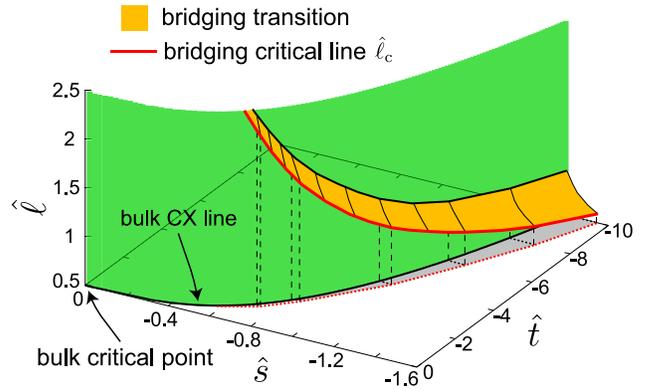}
\caption{\protect  
(Color online) Phase diagram 
in the $\hat{t}$-$\hat{s}$-$\hat{\ell}$ 
space   outside the bulk coexistence  surface (CX) (green), where 
a   surface of a first-order bridging transition (orange) 
is bounded by CX  and  a bridging  critical line (red). 
The critical  line approaches   CX  tangentially at 
$(\hat{t},\hat{s}, \hat{\ell})=(-1,-0.66, 2.6)$.
 } 
\end{center}
\end{figure}

%% 9
\begin{figure}
\begin{center}
\includegraphics[scale=0.75]{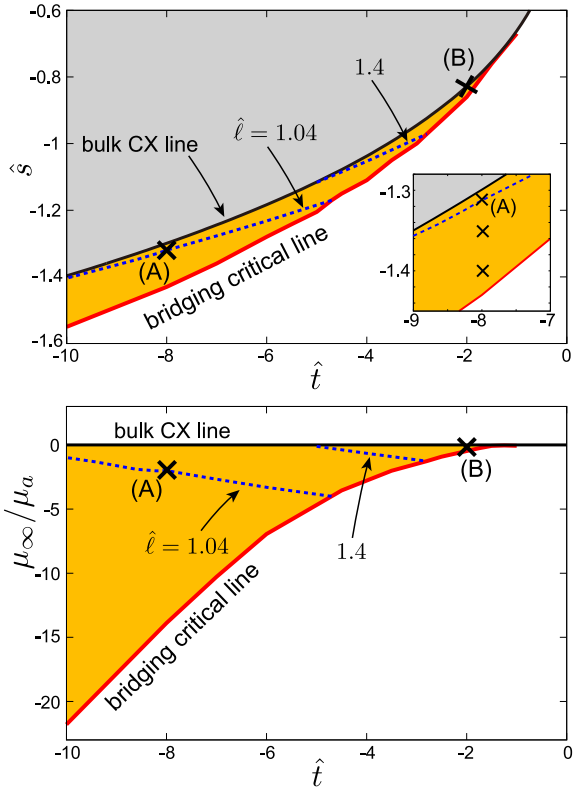}
\caption{\protect  
(Color online) Phase diagrams in the $\hat{t}$-$\hat{s}$ plane (top) 
and  in the $\hat{t}$-$\mu_\infty/\mu_a$ plane (middle), 
where $\mu_a=k_BT_c /a^3 \psi_a$.  A 
bridging transition occurs at some $\hat\ell$ in 
the region between  CX  and the bridging critical line.  
Shown also are cross-sectional  
bridging transition lines at fixed  $\ell=1.4$ and 1.04 
 (blue dotted lines).   In the inset (top),
 a region around $\hat{t}=-8$ is expanded. 
 Particularly for bridging behaviors at  point (A), point (B),  
and  three points $\times$ (inset), 
see  the following figures. } 
\end{center}
\end{figure}

%% 10
\begin{figure}
\begin{center}
\includegraphics[scale=0.8]{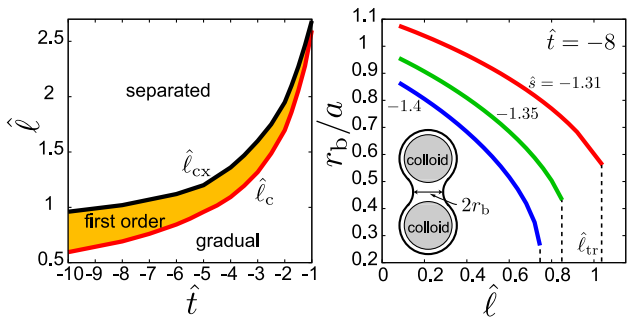}
\caption{\protect  
(Color online)
%%%%%%%%%%%%
Left: Phase diagram 
in the $\hat{t}$-$\hat{\ell}$ plane, where 
 the transition line 
%$\hat{\ell}=  \hat{\ell}_{\rm cx}(\hat{t})$ 
on CX and and the critical line 
%$\hat{\ell}=  \hat{\ell}_{\rm c}(\hat{t})$ 
are written. Separated states are realized for any $\hat s$  above 
the transition line, a first-order bridging transition 
occurs  for some $\hat s$ 
between the two lines, 
and the changeover is continuous 
or gradual for any $\hat s$ below the critical line.  
Right: Bridging radius $r_b$ vs $\hat{t}$ at  
$\hat{t}=-8$ for  $\hat{s}= -1.31, -1.35$, and $-1.4$ with 
$r_b$ being defined in the inset, which increases 
with decreasing $\hat\ell$ and is smaller 
near the critical line. 
 } 
\end{center}
\end{figure}  

%% 11
\begin{figure}
\begin{center}
\includegraphics[scale=0.4]{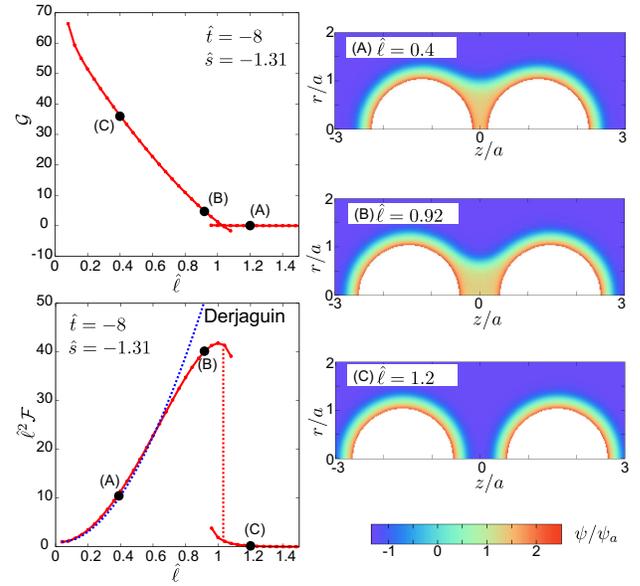}
\caption{\protect  
(Color online) Left: ${\cal G}$ (top) 
and $\hat\ell^2{\cal F}$  (bottom)  vs $\hat\ell$  
for $(\hat{t},\hat{s})=
(-8,-1.31)$ across the bridging transition surface. 
There appear two branches of stationary solutions in a window 
range ($0.96 <\hat{\ell}<1.08$). Maximization of $\cal G$ determines 
the equilibrium state. For $\cal F$, a curve from 
the Derjaguin approximation  (blue dotted line) nicely 
agrees with that from Eq.(3.24). 
Right: $\psi(r,z)/\psi_a$ 
in the $z$-$r$ plane ($r= (x^2+y^2)^{1/2}$) with the same 
 $(\hat{t},\hat{s})$ according to the color bar,  
 where   $({\hat\ell},\psi(0,0)/\psi_a)= $  
(A) $(0.4,1.48)$, (B) $(0.92,1.28)$, and (C) $(1.2,-1.23)$ from above.  
The corresponding points are marked  in the left.
}
\end{center}
\end{figure}

\section{Bridging transition 
between  two colloidal particles}
\setcounter{equation}{0}

%Aomega_loc'ðbridgingŠE–Ê'Ì'Æ'±'ë(\ell/a >0.6)'Å
%Ï•ª'µ'½'à'̂ƁAŠE–Ê'£—Í'ð"äŠr%'µ'ÄŒ©'Ü'µ'½B
%\hat{t}=-8'Å'Í 2 \pi \times a^2 \sigma / k_BT_c = 57.87
%'»'µ'āAomega'ðÏ•ª'µ'½'à'Ì (a/k_BT_c) 2\pi \int r \omega_{loc} dr '́A
%\ell/a=0.08'Å'Í76.59,
%0.2'Å'Í72.80,
%0.4'Å'Í'U'UD'S'V
%bridge 'µ'½Žž'́A
%force\p \Omega/\ell  \sim \pi a \sigma \sim kT (a/ \xi )^2 :
%}9'Ì{\cal G}'ÌŒX'«(=-{\cal F})'́Atransition point'ŁA-37.8Aˆê•û\hat{t}=-8'Å
%-2\pi a^2 \sigma /k_BT_c=-57.9'Å'·B
%'Ü'½A}'P'O'Å'Ítransition point'ŁAbridging branch'Ì{\cal G}'ÌŒX'«'Íb'̋Ȑü'Å-%36.4, c'̋Ȑü'Å'Í-26'Å'·BŠE–ʈʒu'ª³Šm'É"¼Œa'ƈê'v'µ'Ä'¢'é–ó'Å'Í'È'¢'̂ŁA'±%%'Ì'ö"x'Ì'¸'ê'Í' 'Á'Ä'àŽd•û'ª–³'¢'Æ'¢'¤'Æ'±'ë'Å'Í'È'¢'Å'µ'傤'©

In this section, 
 we  study  the 
bridging transition for $\hat{s}<0$ 
between two  colloidal particles  
in a near-critical binary mixture. 
In our case,  $\psi$ assumes the profiles of $\hat{s}<0$ in Fig.1 
in   separated states,  where the adsorption layer has 
a thickness of order $\xi_{\rm av}$ 
in Eq.(3.19). A  bridging transition 
 can then occur in a wide range of   $\ell$($< 2.6a$)  
under the condition $\xi/a \sim 0.3 |\hat{t}|^{-\nu}<1$. 
In the previous papers  
\cite{Vino,Higashi,Bauer,Yeomans,Evans-Hop}, 
     bridging between  two   spheres 
or between a sphere and a plate were  studied  numerically 
for small separation  $\ell$ (say $\sim  0.2a$)  far  from the criticality.

%12
\begin{figure}
\begin{center}
\includegraphics[scale=0.39]{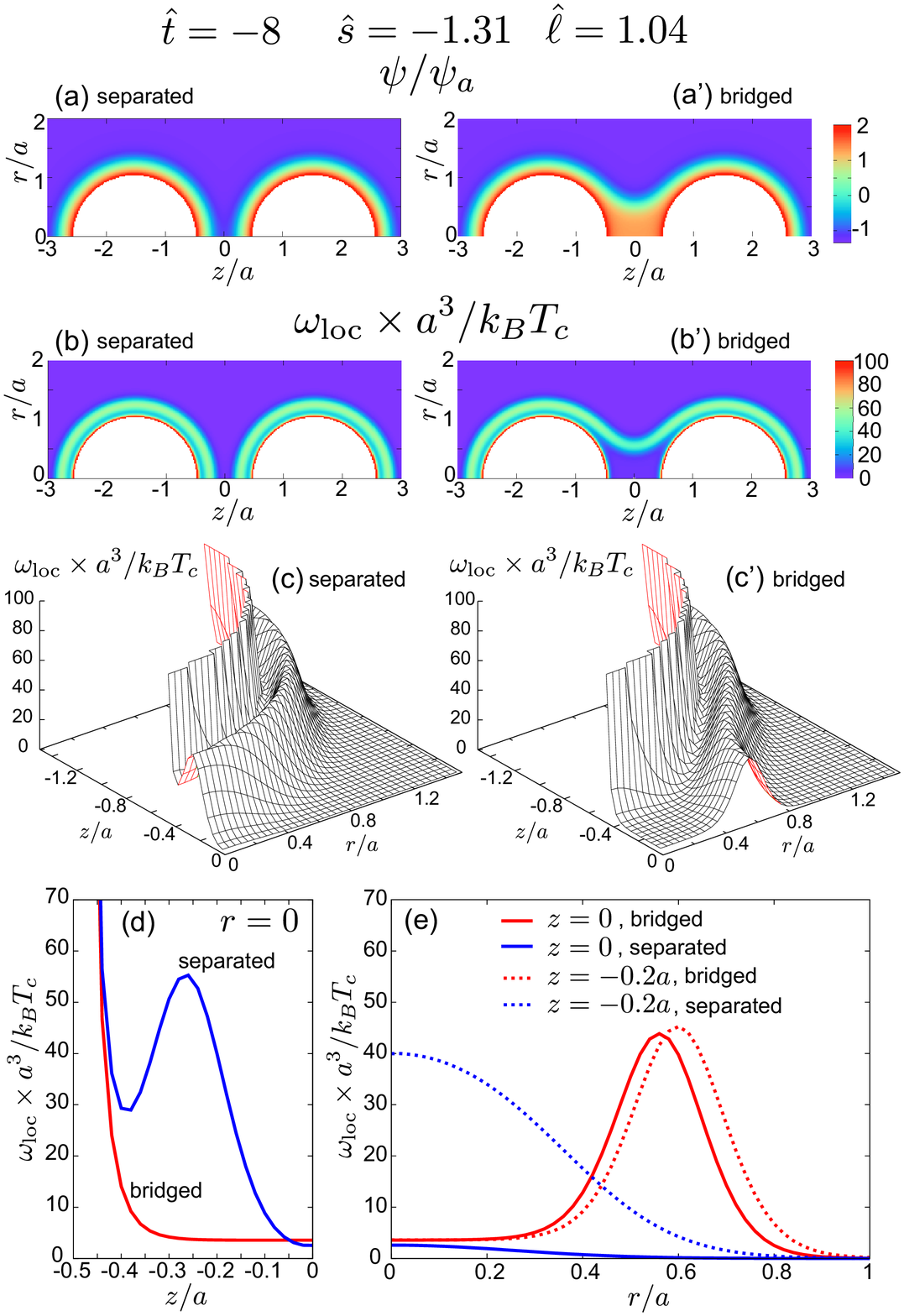}
\caption{\protect 
Separated state (left) and bridged  state (right)  
at a first-order transition point with   $(\hat{t},\hat{s},\hat{\ell})
=(-8,-1.31,1.04)$. Here,   $\psi/\psi_a$ in (a) and (a') 
and $\omega_{\rm loc} a^3/k_BT_c$ in (b) and (b') are 
written in gradation according to the color bars. 
Also  
 $\omega_{\rm loc} a^3/k_BT_c$   is diplayed 
in bird's eye views  for  $z\le 0$ in (c) and 
(c'),  along the $z$ axis on the line $r=0$ in (d), 
and as a function of $r/a$ at  $z=0$ and $-0.2a$ 
in (e). 
}
\end{center}
\end{figure}
%13
\begin{figure}[t]
\begin{center}
\includegraphics[scale=0.39]{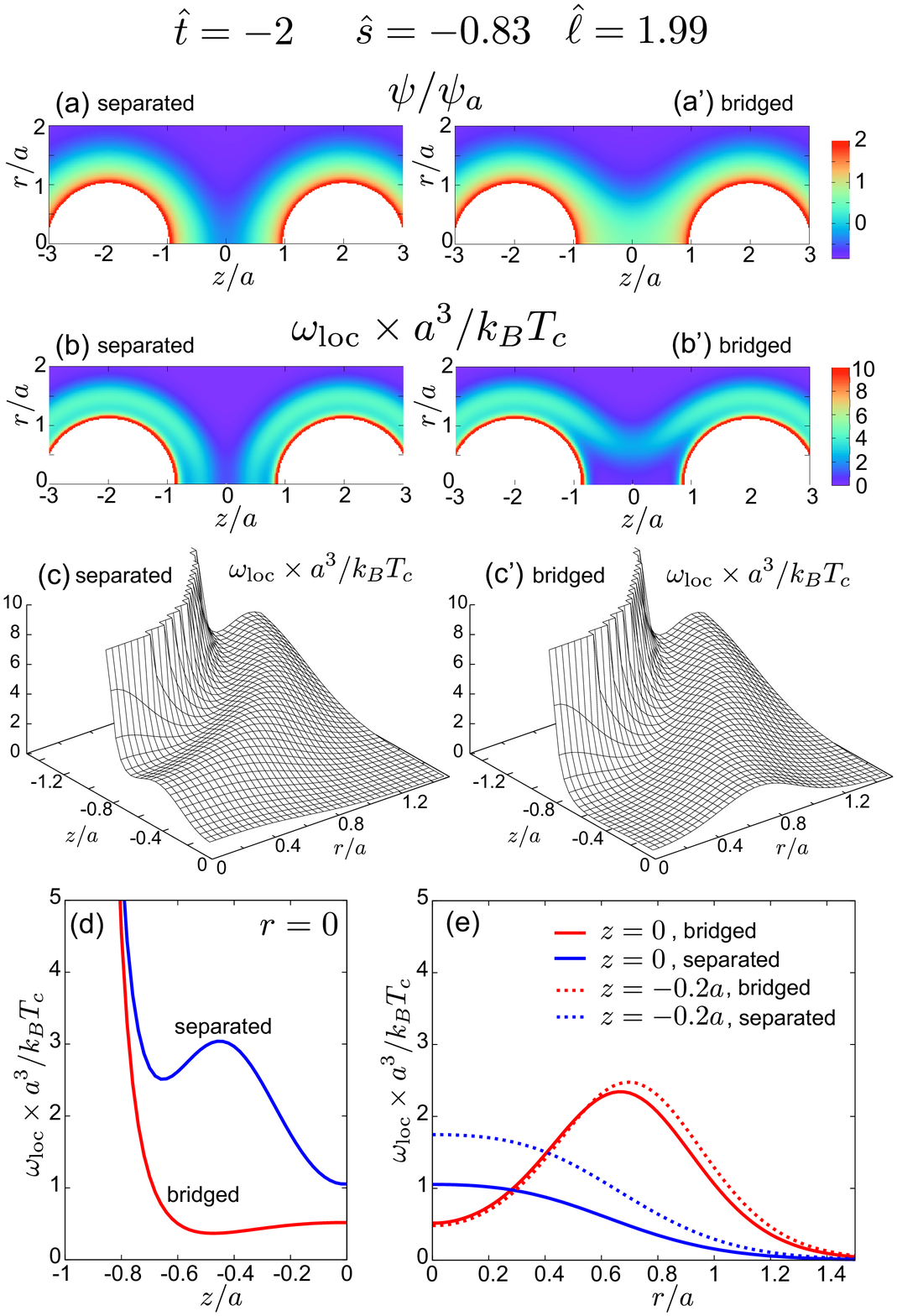}
\caption{\protect
(Color online) 
Separated state (left) and bridged  state (right)  
at a first-order transition point near the bulk criticality 
with  $(\hat{t},\hat{s},\hat{\ell})
=(-2,-0.83,1.99)$, where  
$\psi/\psi_a$ in (a) and (a') 
and $\omega_{\rm loc} a^3/k_BT_c$ in (b) and (b').
Also   $\omega_{\rm loc} a^3/k_BT_c$  
is displayed in bird's eye views in (c) and (c') 
for  $z \le 0$, along the $z$ axis on the 
line $r=0$ in (d), and as a function of $r/a$ at 
 $z=0$ and $-0.2a$ in (e). 
}
\end{center}
\end{figure}

\subsection{Phase diagrams}

In  Fig.8, we first show a  phase diagram 
 of  the bridging transition in the 
$\hat{t}$-$\hat{s}$-$\hat{\ell}$ space outside CX.  
We find  a surface of a first-order bridging  transition,  
 bounded by  CX  and  a bridging  critical line. 
As functions of  $(\hat{t},\hat{s})$, 
the normalized separation $\hat\ell$ may be  written 
on the transition  surface and  on the critical line   as 
\be 
\hat{\ell}= \hat{\ell}_{\rm tr}(\hat{t},\hat{s}),\quad  
\hat{\ell}= \hat{\ell}_{\rm c}(\hat{t}),
\en 
respectively. Across this surface, 
discontinuities appear in  $\cal F$ and the adsorption 
$\Gamma$ in Eq.(3.18), which tend to 
vanish on approaching the critical line. The critical  line 
tangentially ends  on CX at 
$(\hat{t},\hat{s},\hat{\ell})\cong (-1.0,  -0.66, 2.6)$.  
The maximum of $\hat{\ell}$ 
at a transition is thus 2.6.

In Fig.9, phase diagrams in 
the $\hat{t}$-$\hat{s}$ 
and  $\hat{t}$-$\mu_\infty/\mu_a$ planes are presented,  
where $\mu_\infty$ is  related 
to  $\psi_\infty$  by Eq.(3.4) and scaled by 
 $\mu_a=k_BT_c /a^3 \psi_a$.   
In these phase diagrams,  
a first-order bridging transition 
 occurs at some $\hat\ell$ in the region between  
CX  and the bridging critical line, 
%as can be seen in Fig.8, 
where the latter  approaches CX tangentially. 
We also write cross-sectional  
bridging transition lines at fixed  $\ell$ (equal to $1.4$ and $1.04$)  
on the bridging transition surface, 
each starting from CX and ending at 
a point on the critical line.  
In our case, these lines are nearly 
 straight in the two phase  diagrams in Fig.9. 
 Previously, bridging transition lines 
at fixed separation $\ell$ were 
 drawn \cite{Yeomans,Bauer,Gamb}. 
For near-critical films, on the other hand, 
the capillary condensation line 
is detached from CX.   
As  a result, it  is considerably 
 curved in the $\tau$-$\psi_\infty$ plane \cite{OkamotoCasimir}, 
but  is  nearly straight in the 
$\tau$-$\mu_\infty$ plane \cite{Yabunaka}.

The phase behavior at fixed separation $\ell$ is particularly  intriguing. 
In the  left panel of Fig.10, we show a phase diagram 
in the $\hat{t}$-$\hat{\ell}$ plane, where 
we write the critical line $\hat{\ell}= \hat{\ell}_{\rm c}(\hat{t})$ 
and the transition line $\hat{\ell}= \hat{\ell}_{\rm cx}(\hat{t})$ 
on CX. The latter is defined  by 
\be 
\hat{\ell}_{\rm cx}(\hat{t})= 
 \hat{\ell}_{\rm tr}(\hat{t},-{\hat{s}}_{\rm cx}(\hat{t})),  
\en 
where $ -{\hat{s}}_{\rm cx}(\hat{t})
= -0.68 |\hat{t}|^{\beta/\nu}$ is the vallue of 
 $\hat{s}$   on the negative branch of CX. 
These two lines merge at 
 $(\hat{t},\hat{\ell})=(-1, 2.6)$ on CX.
 Then,  let us vary $\hat s$ at fixed $\hat \ell$ and $\hat t$. 
(i) If $\hat{\ell}>\hat{\ell}_{\rm cx}(\hat{t})$, 
separated states are realized without bridging for any $\hat s$. 
(ii) If $\hat{\ell}_{\rm cx}(\hat{t})>\hat{\ell} >\hat{\ell}_{\rm cx}(\hat{t})$,we encounter the transition surface at a  certain $\hat s$ 
  to find a discontinuous change. 
(iii) For   $\hat{\ell} <\hat{\ell}_{\rm cx}(\hat{t})$, 
   a  bridging domain appears 
with a  well-defined  interface 
%(with thickness $\xi_{\rm cx}$)  
close to CX, but disconnection occurs   continuously 
  with incresaing  the distance from CX. 
In this changeover, it is puzzling how the interface 
becomes ill-defined gradually (see Fig.17).

In  the  right panel of Fig.10, 
we plot   the bridging radius 
 $r_b$ vs $\hat \ell$ 
at $\hat{t}=-8$ for $\hat{s}=-1.31, -1.35$, and $-1.4$, 
which correspond to the three marked points 
in the top panel of Fig.9.  
We determine $r_b$ from the condition 
$\psi(r_b,0)=0$ at $z=0$, where $\psi(r,0)$ 
changes from positive to negative at  $r=r_b$ 
 with a bridging domain 
  in the range $\hat{\ell}< 
\hat{\ell}_{\rm tr}(\hat{t},\hat{s})$.  
As a function of  $\ell$ at  each $(\hat{t},\hat{s})$, 
  $r_b$ is shortest at the transition and  
  increases  with decreasing $\hat\ell$. 
It is about  $a$ for sufficiently small $\hat \ell$. 
Also it is  smaller near  the critical line. In fact, 
$r_b\cong 0.2a$ at the transition for  $\hat{s}=-1.4$.

The transition surface is determined 
from minimization of $\Omega$ or  maximization of $\cal G$ from Eq.(3.21). 
In the  left panels of Fig.11, we plot ${\cal G}$ 
and $\hat\ell^2{\cal F}$    vs $\hat\ell$  
for $(\hat{t},\hat{s})=(-8,-1.31)$.    The curve of  $\cal F$ from 
the Derjaguin approximation    nicely 
agrees with that from Eq.(3.24) for $\hat{\ell} \ls 0.6$. 
  For this  ($\hat{t}, \hat{s})$, 
we find  two stationary solutions satisfying Eqs.(3.1) and (3.5) in a 
window range ($0.96 <\hat{\ell}<1.08$ for this example). 
Outside this range,  one  solution 
becomes  unstable and the other one 
remains as  a  stable  solution. 
In the bistable range, $\cal G$ 
is larger on the   equilibrium branch 
and smaller on the  metastable one, 
so  the transition is at the crosspoint of the  two branches of $\cal G$.

 In Fig.11, the  slope of  $\cal G$ is very steep   
with   bridging. 
It is  $ -37.8$ at the transition, 
where  $\hat{\ell}=1.04 \gg \xi/a=0.09$. 
It is further amplified  for smaller $\hat\ell$ 
and is  $-76.6$  at $\hat{\ell}=0.08\cong \xi/a$.   
Here,  for $\ell\gg \xi$,  a well-defined  bridging domain exists 
and  ${\cal G}$ changes with 
a change of its  surface area.  In fact, use of the surface tension 
$\sigma$ in Eq.(2.12) gives  $ 2\pi a^2 \sigma /k_BT_c=57.9$ 
at ${\hat t}=-8$.   Thus,  with a  well-defined bridge, 
 Eq.(3.24) yields the capillary force   \cite{Butt,Butt1},  
\be 
{\cal F} \sim 2\pi a^2 \sigma /k_BT_c\sim (a/\xi)^2 .
\en 
This relation is  valid for  
${\ell}\gs \xi$. For smaller $\ell \ls \xi$, 
the growth ($\sim {\hat\ell}^{-2}$) 
 in  Eq.(3.25) becomes dominant.  
These features will be 
further examined in Figs.12, 13, and 16. 

In the original units, the force with a well-defined bridge   
is of order $k_BT_c a/\xi^2$, which 
 increases as we move 
away  from the bulk criticality. Also in Fig.14 below, 
we shall see that $\cal F$ increases with 
lowering  $\tau$, where bridging occurs continuously. 
However, the exponential tail 
of  the interaction in Eq.(3.27) 
in separated states ($\propto e^{-\ell/\xi}$) 
  increases as the bulk criticality is approached, 
which was  indeed observed experimentally  
\cite{Nature2008}.

In the  right  panels of Fig.11, we 
display  $\psi(r,z)/\psi_a$ 
in the  plane of $r= (x^2+y^2)^{1/2}$ and $z$  
in two bridged and one separated states at different $\hat\ell$.
We can see that the bridging radius $r_b$   is larger 
 in (a) (far below the bridging transition) than in (b) (close to it). 
The midplane between the two particles is 
filled with the phase outside the particles 
in (c) in a separated state.

In these phase diagrams 
 the lowest value of $\hat t$ is $-10$.  
With further loweing $\hat t$, 
there is still a tendency of decreasing 
 $\hat{\ell}_{\rm cx}$  and 
$\hat{\ell}_{\rm c}$.  For example, 
for $\hat{t}=-20$,  we find 
$\hat{\ell}_{\rm cx}=0.69$ on CX and   
$(\hat{\ell}, \hat{s}) = ({\hat{\ell}}_c , {\hat{s}}_c)
= (0.395, -2.0)$  at the corresponding bridging 
critical point.  

%% 14 
\begin{figure}
\begin{center}
\includegraphics[scale=0.45]{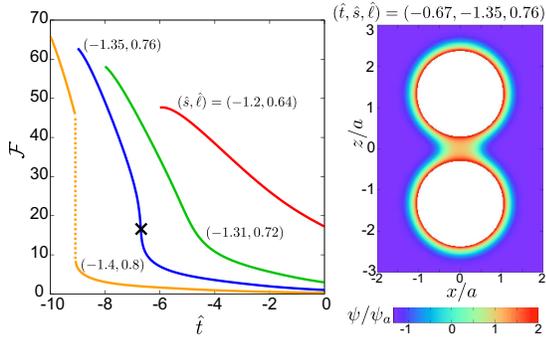}
\caption{\protect  
(Color online) Left: 
 $\cal F$ vs  $\hat{t}$ 
for  $(\hat{s},\hat{\ell})=(-1.2,0.64)$, $(-1.31,0.72)$, 
$(-1.35,0.76)$,and $(-1.4,0.8)$ from right. 
The slope $\p{\cal F}/\p {\hat t}$ becomes  steeper  
near the bridging critical line. On the third curve (from right), 
a critical point ($\times$) is passed. 
On the fourth curve, a  jump due to  a bridging transition 
appears. Right:  $\psi(x,0,z)/\psi_a$ on the $xz$ plane 
according to the colar bar  at a critical point 
$(\hat{t}, \hat{s},\hat{\ell})=(-0.67, -1.35,0.76)$ ($\times$ 
in the left panel). It is equal to 1.08 at the center $x=z=0$. 
}
\end{center}
\end{figure}

\subsection{Profiles at transition and critical points}

In Figs.12 and 13, we  compare the profiles of 
$\psi(r,z)$ and $\omega_{\rm loc}(r,z)$ 
in separated and bridged states 
at  two typical transition points 
on the bridging transition surface 
in Fig.8. That is, 
 $(\hat{t},\hat{s},\hat{\ell})$ 
is $(-8,  -1.31, 1.04)$ in Fig.12  and 
is $(-2,-0.83,1.99)$ in Fig.13. 
These points   correspond to  
points (A) and (B) in Fig.9. 
The former in Fig.12 is relatively far from 
the bulk criticality with $\xi/a \cong 0.09$ 
and the interface is well-defined. 
The latter in Fig.13 
is closer to it with $\xi/a \cong 0.19$ 
and the interface is  broadened 
and the separation 
is widened to  $\hat{\ell}=1.99$.

More remarks on Figs.12 and 13 are as follows.
(i) The profiles 
of  $\psi/\psi_a$  are distinctly different 
 in the separated and bridged states. 
Its  midpoint value  
  is 1.27 in (a) and $ -1.07$ in (a') in Fig.12, while it 
 is 0.54 in (a) and $ -0.31$ in (a') in Fig.13.
(ii) We can see layer  regions with a peak in 
 $\omega_{\rm loc} a^3/k_BT_c$, which  
enclose the colloid surfaces  at a distance  of order 
$\xi_{\rm ad}$ in Eq.(3.19) except for the bridged surface regions. 
See Fig.1 for the  profile of ${\hat s}=-1.31$ 
around  a  single particle. 
These layers around the two spheres 
   are separated in (b), while 
 they  are detached from the colloid surfaces 
in the bridged parts in  (b'). 
(iii) We also  display $\omega_{\rm loc} a^3/k_BT_c$   
in bird's eye views in 
(c) and (c'), Comparing them, we recognize 
 how a discontinuous change 
occurs with the total grand potential unchanged. 
(iv) In the bottom plates, we plot     one-dimensional profiles of  
$\omega_{\rm loc} a^3/k_BT_c$ in the two states. 
They are presented   
along the $z$ axis at  $r=0$ in (d)   
and   along the $r$ axis  at  $z=0$ and $-0.2a$ 
in (e).  Note that the integral 
$ \int_0^\infty dr r \omega_{\rm loc} a/k_BT_c$ 
is equal to ${\cal F}/2\pi$ from Eq.(3.24) 
and is of order $r_b a\sigma$ 
with $\sigma$ being the surface tension.

 Figsures 12 and 13  demonstrate   that   
 there should be  a balance between 
the free energy cost of creating 
a bridge  ($\sim \pi \sigma  \ell r_b$) 
and the free  energy decrease 
on the colloid surfaces ($\sim - \pi \sigma r_b^2$) 
 at the transition (see (d) and (e)). 
The origin of the latter is evident  
from comparison of the 
two curves of ${\hat s}= \pm 1.31$ in   Fig.1. 
Also in  Fig.13, 
we have  ${\cal F}=6.79$ 
with bridging and  
 $2\pi  a^2 \sigma /k_BT_c =10.09$, in agreement with Eq.(4.3). 
The corresponding values in Fig.12 have already 
been given above Eq.(4.3). 
%e have written 
% $ {\cal F}= -37.8$ with bridging at the transition 
% and  $ 2\pi a^2 \sigma /k_BT_c=57.9$ for 

We  also examine  
the behavior of $\cal F$ and the 
profile of $\psi$ near  the critical line. 
In Fig.14, we plot  $\cal F$ vs  $\hat{t}$ 
for four sets of  $(\hat{s},\hat{\ell})$.  Here, the 
force $\cal F$ increases with lowering $\hat{t} <0$. 
The right two curves are   in  regions with  
 $\ell<\ell_{\rm c}$ in Fig.8  
and there is no discontinuous change (as in Fig.17 below).  The third 
 curve  meets  a critical point 
$(\hat{t}, \hat{s},\hat{\ell})=(-0.67, -1.35,0.76)$.
On these curves, bridging is  achieved  continuously 
as $\hat t$ is lowered. 
The fourth   curve passes through   
the bridging transition surface 
and exhibits a discontinuous change.  
The slope $\p{\cal F}/\p {\hat t}$ 
becomes  steep near the  critical line 
and diverges on it.

%% 15 
\begin{figure}
\begin{center}
\includegraphics[scale=0.48]{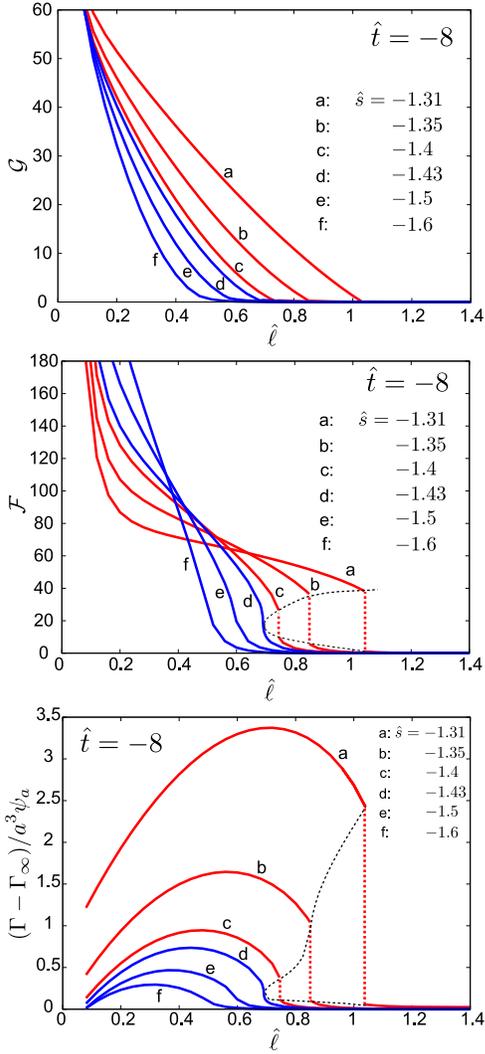}
\caption{\protect  
(Color online)  ${\cal G}$ (top),  
$\cal F$ (middle), and  normalized 
excess adsorption $(\Gamma-\Gamma_\infty)/a^3\psi_a$ (bottom) 
vs $\hat\ell$    
for  $\hat{t}=-8$, where $\hat{s}$ is (a)$-1.31$, (b)$-1.35$, (c)$-1.4$, 
(d)$-1.43$, (e)$-1.5$, and (f)$-1.6$. 
For (a), (b), and (c), 
 a discontinuous bridging transition occurs  at $\hat\ell=1.04$, 
$0.85$ and $0.745$, respectively.
 For (d), (e), and (f), the curves are 
 continuous in the whole range of $\hat\ell$, where   
$\Gamma-\Gamma_0$ becomes negative 
for $\hat{\ell}<0.08$ (not shown).
For (d), the curve nearly passes 
through a critical point, where    
$-\p{\cal F}/\p {\hat\ell}$ and $- (\p \Gamma/\p \hat{\ell})/a^3\psi_a  
$ are very large. }
\end{center}
\end{figure}  

\subsection{Overall behaviors}

In Fig.15, we plot 
${\cal G}$, $\cal F$, and  the normalized 
excess adsorption $(\Gamma-\Gamma_\infty)/a^3\psi_a$ 
vs $\hat\ell$ at $\hat{t}=-8$ 
for six values of  $\hat{s}$. Here, 
$\Gamma$ is the adsorption in Eq.(3.18) 
and $\Gamma_\infty$ is its value  for large separation. 
A discontinuous bridging transition occurs 
for $\hat{s}= -1.31$, $-1.35$, and $-1.4$,
while there is no discontinuity for
$\hat{s}= -1.43$, $-1.5$, and $-1.6$. 
 In the latter, 
$\Gamma-\Gamma_0$ becomes negative 
for $\hat{\ell}<0.08$ (not shown).
For (d), the curves nearly pass  
through a critical point, where   we have 
very steep slopes:  
$\p{\cal F}/\p {\hat\ell}= -3413$ and 
$ (\p \Gamma/\p \hat{\ell})/a^3\psi_a  
= -39.5$. This behavior indicates 
divergence of   $\p {\cal F}/\p \hat{\ell}$ and 
$\p { \Gamma}/\p \hat{\ell}$ on the bridging critical line. 
In addition, the curves of  $(\Gamma-\Gamma_\infty)/a^3\psi_a$ 
vs $\hat\ell$ exhibit rounded maxima at an intermediate 
$\hat \ell$.  This  behavior can be understood from 
the right panel of Fig.10, where 
 the bridging radius $r_b$ increases   
with decreasing $\hat\ell$.

In Fig.15, the formula (4.3) 
for bridged states holds for $\hat{\ell} \gs \xi/a= 0.09$. 
For smaller $\hat \ell$, $\cal F$ diverges as in Eq.(3.25). 
In Fig.16, we thus plot 
 $\psi(r,0)/{\psi_a}$  and $r 
\omega_{\rm loc}(r,0)\times  2\pi a^2/k_BT_c$  
as functions of $r/a$ at $z=0$ 
in the midplane at   $(\hat{t}, \hat{s})=(-8, -1.31)$ 
for $\hat{\ell}=0.4, 0.2$, and 0.08. 
We recognize growing of 
$\psi$ and $\omega_{\rm loc}$ on the midplane 
with decreasing ${\hat\ell}\ls \xi/a$.

We also examine how a 
 continuous changeover between bridged and separated states 
is achieved  for small $\hat{\ell}<\hat{\ell}_c(\hat{t})$. 
This case   has  been 
mentioned in the explanation of Fig.10.  
In Fig.17, we display  
 $\psi(r,0)/{\psi_a}$  and $r 
\omega_{\rm loc}(r,0)\times  2\pi a^2/k_BT_c$  
vs  $r/a$ at $z=0$ for  $\hat{t}=-8$ and    $\hat{\ell}=0.5$. 
Here, $\hat s$ is decreased from a value close to CX, 
$-1.31$, to smaller values away  from CX. 
The  profile of  $\psi(r,0)/{\psi_a}$  at  
$\hat{s}= -1.31$ indicates   the presence of 
 a well-defined interface  
with a  thickness of order $\xi$. However, 
with decreasing $\hat s$  from -1.31 to -1.7, 
the profile of  $\psi(r,0)/{\psi_a}$ is gradually broadened 
and $\cal F$ decreases from  $67.7$ to 13.8.

%% 16 
\begin{figure}
\begin{center}
\includegraphics[scale=0.43]{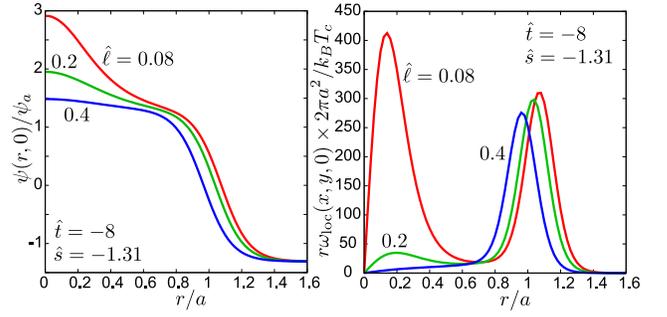}
\caption{\protect  
(Color online) 
Midplane profiles: 
 $\psi(r,0)/{\psi_a}$ (left) and $r 
\omega_{\rm loc}(r,0)\times  2\pi a^2/k_BT_c$ (right) vs $r/a$  
at $z=0$ 
%($r=(x^2+y^2)^{1/2})$ 
 for $(\hat{t},\hat{s})=(-8,-1.31)$. Area below each curve 
(right)  gives  $\cal F$. Here,   $\hat{\ell}=0.4$, 
0.2, and $0.08$, 
for which  ${\cal F}=71.7$, 87.4, and 185,  respectively.
   At $r \sim a$,  a well-defined interface exists. 
In the center region, $\psi$ and  $\omega_{\rm loc}$ 
grow for ${\hat \ell} \ls \xi/a=0.09$.     
}
\end{center}
\end{figure}  
%% 17 
\begin{figure}
\begin{center}
\includegraphics[scale=0.43]{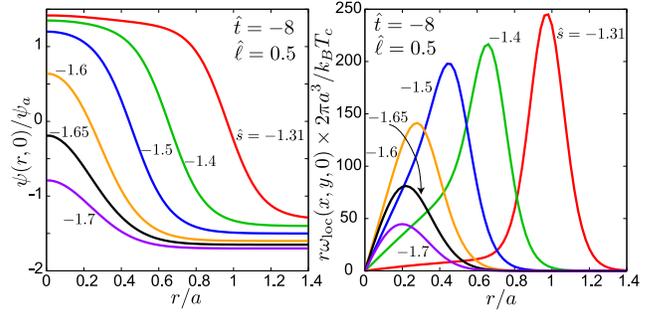}
\caption{\protect  
(Color online) 
Midplane profiles: 
 $\psi(r,0)/{\psi_a}$ (left) and $r 
\omega_{\rm loc}(r,0)\times  2\pi a^2/k_BT_c$ (right)  
vs $r/a$  at $z=0$ 
%($r=(x^2+y^2)^{1/2})$ 
 for $(\hat{t},\hat{\ell})=(-8,0.5)$. Area below each curve 
(right)  gives  $\cal F$.  
 Here,  $\hat{s}$ is varied  as $-1.31, -1.4, -1.5, -1.6$, -1.65, and $-1.7$, 
for which  $\cal F$ is equal to 
 $67.7$, 76.8, 65.4, 40.1,  25.8,  and 13.8, respectively.  
In this case, $\hat{\ell}< \hat{\ell}_c$ holds, so 
changeover between bridged and  separated states is 
continuous.   
}
\end{center}
\end{figure}  

\subsection{Stability of  separation distance}

 In Figs.14 and 15, the derivatives   
$\p {\cal F}/\p {\hat\ell} (\propto 
\p^2\Omega/\p\ell^2)$ and 
$\p {\cal F}/\p {\hat t} (\propto 
\p^2\Omega/\p\ell\p{\tau})$ are negative 
and tend to  diverge  as the bridging critical 
line is approached.  
We mention one implication of this 
singular behavior.

In this paper, we have been fixing 
the colloid separation 
$\ell$ at a constant. To achieve this constraint, 
let us suppose the presence of an externally 
applied potential $U_{\rm ext}(\ell)$ 
between  two colloidal particles. 
It is worth noting that optical tweezers 
have been used to trap colloidal particles 
at small separation \cite{Nature2008,Kimura}. 
In equilibrium,  we should minimize the sum 
$\Omega+ U_{\rm ext}$ with respec to $\ell$. Then, the 
equilibrium separation  $\ell$  is determined from  
\be 
F_{\rm ext}= -\frac{\p}{\p \ell}U_{\rm ext}=
\frac{1}{a}k_BT_c  {\cal F} .  
\en  
In order to ensure the stability of 
 this equilibrium separation $\ell$, 
we need to require 
\be 
K_{\rm ext}= \frac{\p^2}{\p \ell^2}U_{\rm ext}
> \frac{1}{a^2}k_BT_c 
\bigg(- \frac{\p{\cal F}}{\p \hat{\ell}} \bigg) 
\en 
where $K_{\rm ext}(\ell) 
$ is the spring constant of the 
externally applied potential. 
The  thermal fluctuation of 
the separation $\ell$ is increased 
with decreasing the effective 
spring constant $K_{\rm eff}= 
K_{\rm ext} 
+ (k_BT_c/a^2) \p {\cal F}/\p {\hat\ell}$ while  $K_{\rm eff}>0$.  
However, there is a possibility 
of  violation of the  inequality (4.5) 
or negativity of $K_{\rm eff}$ 
sufficiently close 
to the bridging critical line, where 
the colloid configuration determined  from Eq.(4.4) 
is unstable.

\section{Summary and remarks}

We have investigated  
  the adsorption-induced  interaction between two 
neutral colloidal particles with common radius $a$ 
  in a near-critical binary mixture. 
 Use has been made of our  local functional theory \cite{OkamotoCasimir}. 
In the strong adsorption limit, we have calculated  the normalized 
 free energy deviation $\cal G$ (with minus sign)  
and  the normalized force  $\cal F$   
as universal  functions of 
scaled reduced temperature $\hat{t}= \tau/\tau_a$ (where 
$\tau_a= (\xi_0/a)^{1/\nu})$, scaled 
reservoir order parameter 
$\hat{s}= \psi_\infty/\psi_a$ (where $\psi_a \sim \tau_a^\beta$), and 
scaled separation distance 
$\hat{\ell}=\ell/a$.  

 Main results are as follows.\\ 
(i) We have expressed  the forces  
for many colloidal particles  in  Eq.(3.6)  and the force  
between  two neutral 
colloidal  particles in Eq.(3.24) using  
the stress tensor due to  the order parameter deviation. 
Some general discussions on this aspect are given in Appendix A.  
Generalization including charges will be presented 
in another paper.\\
(ii) The  interaction is much enhanced  for $\hat{s}<0$ 
as in Figs.3-7, where 
 the component favored by the colloid surfaces is   poor in the reservoir 
and the order parameter disturbances around the surfaces are large 
as  in Fig.1. It  is $10$-$100$ times larger than 
at the bulk criticality. \\
(iii) The   Derjaguin approximation   \cite{Is,Butt} 
can be made on the force $\cal F$ for  $\hat{\ell} \ls  1$ 
on the basis of the results for films in our previous paper 
\cite{OkamotoCasimir}, as discussed in Appendix B. It 
cannot describe the bridging transition, 
but it predicts the short separation growth 
in Eq.(3.26) and the exponential decay   
for large $\hat{\ell}$ in Eq.(3.27). 
These  results agree with the calculations  
  from Eq.(3.24) for  $\hat{\ell} \ls  1$.\\
(iv) We  have compared    
  the van der Waals interaction and 
the   adsorption-induced interactions.  
The former  may be neglected 
at off-critical compositions 
and particularly at a bridging transition 
even for typical values of the Hamaker constant 
$A_{\rm H}(\sim 10^{-19}$J), as shown in Fig.7.\\    
(v) We have found  a  surface of 
a first-order bridging transition $\hat{\ell}= 
\hat{\ell}_{\rm cx}(\hat{t},\hat{s})$ 
 in the $\hat{t}$-$\hat{s}$-$\hat\ell$ space in Fig.8, 
across  which  a discontinuous change occurs    between 
separated and bridged states. This surface starts from the bulk coexistence 
surface and  ends at  a bridging critical line $\hat{\ell}= 
\hat{\ell}_{\rm c}(\hat{t})$. 
The  discontinuity vanishes and the derivatives of the force 
with respect to $T$ and $\ell$ diverge 
 on   the critical line as in Figs.14 and 15.  
  The critical separation  ${\hat{\ell}}_c$ 
decreases with decreasing $\hat t$, 
which assumes the maximum  2.6 at ${\hat t}=-1.0$ 
and is  $0.395$ at ${\hat t}=-20$.
\\
(vi) We have calculated $\cal G$, $\cal F$, and 
the excess adsorption $\Gamma-\Gamma_\infty$ 
for various parameters in Fig.15. 
With a well-defined bridging domain with $\ell \gs \xi$, 
${\cal F}$ is given by 
the capillary force  proportional to the surface tension 
$\sigma$ in Eq.(4.3).  For $\ell \ls \xi$, 
${\cal F}$ grows as $\hat{\ell}^{-2}$ in accord with 
the de Gennes-Fisher theory. \\
(vii) We have changed  $\hat s$  (or $\hat t$) 
 away from  the  bulk coexistence 
surface fixing  $\hat\ell$  
below $\hat{\ell}_{\rm c}$ in Subsec.IVC. 
There, we have found continuous changeover between 
bridged and  separated states as in Figs.14 and 16.\\ 
(viii) 
We have pointed out a possibility 
of  an instability of the colloid separation 
distance near  the bridging critical line 
where  $\p {\cal F}/\p {\hat \ell}$ 
diverges.\\

We give some remarks below.\\ 
(1) To measure the force between colloidal particles, 
the geometry of  a sphere  
and a plate has mostly been used 
 \cite{Nature2008,Higashi}, 
while the geometry of two spheres 
was also used in  a liquid crystal solvent \cite{Kimura}.   
In these two geometries,  
we expect essentially the 
same theoretical results  for 
near-critical fluids. Systematic experiments 
on the force and the bridge formation  
at off-critical compositions near the bulk criticality 
should  be promising.\\ 
(2) There can arise repulsion between 
 solid objects  with  asymmetric   
boundary conditions (with different signs of  $h_1$) 
 \cite{Law,Gamb,Nellen}. The adsorption-induced 
interaction in such asymmetric conditions 
 should also be studied.\\  
(3) Real colloidal particles 
are usually charged  and  the charge effect 
can be crucial \cite{Beysens,Maher,Nature2008,Bonn,Guo}. 
For example, between a sphere  and a plate, 
Hertlein {\it et al.}\cite{Nature2008} measured 
the adsorption-induced attractive 
interaction for $\ell \gs 0.1 \mu$m 
with $a=1.85\mu$m for various $\tau$ at the critical composition. 
In their experiment, 
the screened Coulomb interaction was   
dominant for smaller $\ell$ and 
decayed exponentially $(\propto e^{-\kappa \ell})$ 
with salt, where the screening length 
$\kappa^{-1} (= 12$nm) was shorter than 
$\ell$ measured. 
\\
(4)  The degree of ionization 
 depends on the  composition and  
the  ion  densities. In aqueous fluids, 
the colloid surface can be hydophobic 
for  weak ionization and  hydrophilic with progress of 
ionization \cite{Okamoto,Maher,Beysens}.  
Futhermore, added  salts 
act as selective impurities to cause precipitation 
forming a wetting layer on the surfaces \cite{Onukireview,Okamoto}. 
Aggregation of colloids    depends  
on these elements. \\ 
(5)  The colloidal particles interact with the two components 
  differently in  a mixture solvent. They constitute 
a three component system, where 
the phase separation behavior  is greatly 
altered  by a small amount of 
the colloidal particles acting as selective impurities 
 \cite{Onukireview,Kaler,Maher}. \\
(6) Dynamics of bridging and aggregation 
of colloidal particles should be of great interest, 
where the hydrodynamic flow is 
crucial\cite{Yabunaka,Teshi}. Dynamical aspects 
have not yet been fully  studied 
experimentally. Simulations 
on the dynamics  of charged colloids, 
 is   complicated, 
where we need to integrate  the dynamic equations 
for the composition,  the ions,  and  the collodal particles  
\cite{Yabunaka,Onukibook}. \\
(7) We should  examine the nanobubble 
bridging in water \cite{bubble}. 
From our viewpoint, nanobubbles can appear with addition of a small 
amount of hydrophobic impurities in water 
 \cite{Onukireview}. 
  Particularly intriguing is 
 dynamics of bubble 
formation and disruption  
upon a  pressure change  \cite{Teshi}.  
\\

\begin{acknowledgments}
We would like to thank Daniel Beysens 
%and Daniel Bonn 
for valuable discussions, 
 This work was supported by Grant-in-Aid 
for Scientific Research  from the Ministry of Education, 
Culture,  Sports, Science and Technology of Japan.  
\end{acknowledgments}

\vspace{2mm}
\noindent{\bf Appendix A: 
Adsorption-induced force between colloidal particles  
in terms of stress tensor}\\
\setcounter{equation}{0}
\renewcommand{\theequation}{A\arabic{equation}}

We consider two configurations 
 of colloid particles in a near-critical fluid. 
That is, the colloid centers are at 
 ${\bi R}_\alpha$ in one configuration  
and at ${\bi R}_\alpha +\delta 
{\bi R}_\alpha$ in another slightly displaced 
one  ($\alpha=1, 2,\cdots$). 
In these two states, we write the   profiles of $\psi$
 as $\psi({\bi r}) $ and $\psi' ({\bi r}')$ using different symbols. 
The space  positions  are written as  
  $\bi r$ and   ${\bi r}'$, respectively. 
We are interested in the difference 
between  the grand potentials,  
 $\Omega= \Omega( \{{\bi R}_\alpha\})$ and   
 $\Omega' = \Omega( \{{\bi R}_\alpha+\delta{\bi R}_\alpha\})$, 
for  these  two states.  
From Eqs.(3.2) and (3.3),  the grand potential 
$\Omega'$ for $\psi'({\bi r}')$  is written as  
\be 
 \frac{\Omega'}{k_BT_c} 
 = \int'\hspace{-1mm}  d{\bi r}' [ {\hat\omega}(\psi') 
+ \frac{C(\psi')}{2}|\nabla'\psi'|^2 ]  - \int \hspace{-1mm} 
dS' h_1 \psi' ,
\en  
where  $\psi'=\psi'({\bi r}')$, $\nabla'= \p/\p {\bi r}'$, 
  and 
\be 
\hat{\omega}(\psi) = 
[f(\psi)-f_\infty - \mu_\infty (\psi-\psi_\infty)]/k_BT_c.
\en 
The $\int' d{\bi r}'$ 
is the  integral  outside the displaced 
colloidal particle, while  $\int dS'$ is that  on their spherical surfaces. 
We assume $\psi\to \psi_\infty$ far from the 
colloidal particles.  

As a mathematical technique, we assume a mapping relation between the positions ${\bi r}'$ and  ${\bi r}$  as 
\be 
{\bi r}'= {\bi r}+ {\bi u}({\bi r}), 
\en 
where ${\bi u}$  is a  {\it displacement} vector 
vanishing  far from  the colloidal particles. 
Its surface value on the $\alpha$-th colloid particle is 
given by  $\delta{\bi R}_\alpha$.  
We rewrite the right hand side of Eq.(A.1) 
by changing  ${\bi r}'=(x', y',z')=(x'_1, x'_2,x'_3)$ 
to ${\bi r} =(x, y,z)=(x_1, x_2,x_3)$. To   first order in $\bi u$, 
we may set  $d{\bi r}'=d{\bi r}(1+ \nabla\cdot{\bi u})$ 
and $\p/\p x_i' = \p/\p x_i - \sum_j D_{ij} \p/\p x_j$, 
 where ${D_{ij}} $ is the strain tensor, 
\be   
 D_{ij}= \p u_j/\p x_i .
\en 
The deviation of the order parameter is written as  
\be 
\delta\psi ({\bi r}) = \psi'({\bi r}')- \psi({\bi r}). 
\en 
To   first order in $\bi u$ and $\delta\psi$, 
we  calculate $\delta\Omega=\Omega'-\Omega$ as  
\bea 
&&\hspace{-5mm} \frac{\delta\Omega}{k_BT_c} 
= \int d{\bi r}\bigg [(
\hat{\omega}+   \frac{C}{2} |\nabla \psi|^2)\nabla\cdot{\bi u} 
+ (\frac{\p \hat{\omega}}{\p \psi}  
+    \frac{ C'}{2} 
{|\nabla \psi|^2} )\delta\psi  \nonumber \\
&&\hspace{-4mm} + C\nabla\psi\cdot \nabla\delta\psi 
-C \sum_{ij}  D_{ij} \nabla_i\psi\nabla_j\psi \bigg] - \int dS h_1\delta\psi , 
\ena 
where $C'= \p C/\p \phi$, 
 $\nabla_i= \p/\p x_i$,  and $\int d{\bi r}$ is the  integral  outside the  
colloidal particles  at the original colloid positions, 
 and $\int dS$ is that  on their surfaces. 
Using    $\Pi_{\psi ij} $  in Eq.(3.7), 
  $\Pi_{\infty}$ in  Eq.(3.9), and  
$\delta\Omega/\delta\psi=\delta F_b/\delta\psi- 
\mu_\infty$ (see Eq.(3.5)), we simplify 
Eq.(A6) as   
\bea 
&&\hspace{-5mm}
\delta\Omega= \int d{\bi r} \sum_{ij}
(\Pi_{\infty}\delta_{ij}-\Pi_{\psi ij})  D_{ij} 
+ \int d{\bi r} \frac{\delta\Omega}{\delta\psi}\delta\psi 
\nonumber\\
&& -k_BT_c \int dS[ C {\bi n}\cdot\nabla\psi+h_1]\delta\psi. 
\ena 
This  relation is general and  valid even in nonequilibrium.

In this paper, we assume that the original state is in equilibrium. 
That is, we assume the equilibrium relations (3.1) and (3.5) 
for $\psi({\bi r})$. Then, only the first term remains in Eq.(A7). 
Further using  the equilibrium relation 
  $\sum_j \nabla_j \Pi_{\psi ij} =0$ 
 in the fluid and 
${\bi u} = \delta{\bi R}_{\alpha }$ 
on the surface of the $\alpha$-th 
 colloidal particle,   we may rewrite Eq.(A7) as 
\be
\delta\Omega  = \sum_\alpha 
\int_\alpha   dS  \sum _{ij} 
(\Pi_{ij}-p_{\infty}\delta_{ij}) n_{\alpha i}
 \delta R_{\alpha j},
\en
where $\int_\alpha   dS $ 
is the integral on the surface 
of the $\alpha$-th colloidal particle  and 
${\bi n}_\alpha = (n_{\alpha x}, n_{\alpha y}, 
n_{\alpha z})$ 
is the normal unit vector. This yields  Eq.(3.6). 

In the equilibrium case 
of  two colloidal particles  in Fig.2,  we set $\delta {\bi R}_1= 
\delta\ell {\bi e}_z$ 
and  $\delta {\bi R}_1= 0$, where  ${\bi e}_z$ is the unit vector 
along the $z$ axis. From  Eq.(A8), $\p\Omega/\p\ell=\lim_{\delta\ell\to 0}
\delta\Omega/\delta\ell$  is obtained  as  
\be
\frac{\p \Omega}{\p \ell}   = 
\int_1  dS  \sum _{i} 
(\Pi_{iz}-\Pi_{\infty}\delta_{iz}) n_{1 i} .
\en
where the surface integral is on the surface of 
the first colloidal particle. 
However, the above formula 
is not suitable  for  numerical calculations 
in the strong adsorption case. 
To devise a more  convenient 
one, we integrate  the  
equilibrium equation $\sum_j 
\nabla_j{\Pi}_{zj}  =0$ 
in the fluid region 
bounded by the upper colloid surface, 
 a  semisphere surface $S_{\rm semi}$, and 
 a circular surface $S_{\rm mid}$  (see Fig.1), 
where the latter surfaces are represented by 
\bea 
&&\hspace{-5mm} S_{\rm semi} =\{(x,y,z) | z>0, x^2+y^2+z^2=L^2 \},
\nonumber\\ 
&&\hspace{-5mm}
 S_{\rm mid} =\{(x,y,z) | z=0, x^2+y^2=L^2 \}.
\ena 
Then $\p\Omega/\p\ell$ in Eq.(A9) 
is equal to the sum of the surface integrals on 
$S_{\rm mid}$ and $S_{\rm semi}$. 
In the limit of large $L$,  
the integral  on $S_{\rm semi}$  vanishes 
and that on $S_{\rm mid}$ yields  
 Eq.(3.24), where we use  Eq.(3.10) 
and the relation 
$\p\psi/\p z=0$ on the midplane.

\vspace{2mm}
\noindent{\bf Appendix B: 
Derjaguin approximation }\\
\setcounter{equation}{0}
\renewcommand{\theequation}{B\arabic{equation}}

In  non-bridging situations (with $ \xi \ll a$),  
we may use  the   Derjaguin approximation  
  for $\ell \ls  a$ \cite{Is,Butt}. 
Though not exact, it 
provides a simple relation between 
 the interaction free energy  between two 
colloidal particles  
and  that  between two plates  in the common boundary 
conditions. 
It is justified  when the two spheres are 
 closely   separated without formation of a  bridging domain. 

We write 
the grand potential for a film  per unit area 
  as  $ \Omega_{\rm f}(D)$, 
 where $D$ is the film thickness.
Then,  we have  
\be
\Omega (\ell) -\Omega_\infty 
\cong \pi a\int _\ell ^\infty dD 
[ \Omega_{\rm f} (D)-\Omega_{{\rm f}\infty}],
\en 
where $\Omega_{{\rm f}\infty}$ is 
the limit of $\Omega_{\rm f}$  for large $D$. 
 Here, we have set  $D= \ell+ r^2/a$ to change 
the integration on the $xy$ plane as  $\int dxdy= 2\pi \int dr r= 
\pi a \int dD$, as  in Eq.(3.25).  
For Ising-like near-critical systems, 
$ \Omega_{\rm f} (D)$  is expressed in 
the de Gennes-Fisher scaling form \cite{Fisher,Fisher-Yang} as 
\be
 \Omega_{\rm f} (D) = \Omega_{{\rm f} \infty}
 -k_BT_c D^{-2}\Delta (t, s),
\en
in three dimensions. In our previous paper \cite{OkamotoCasimir},
 we calculated  $\Delta (t, s)$ in the strong  adsorption limit 
as a universal function of 
two scaling parameters  $t$ and $s$ defined by  
\bea
&& t=\tau (D/\xi_0)^{1/\nu}, \nonumber\\
&& s=\psi_\infty /\psi_D, 
\ena
where  $\psi_D  =
  1.47 b_{\rm cx}(\xi_0/D)^{\beta/\nu}$. 
If $D$ is replaced by $a$, we obtain $\hat t$ and $\hat s$  
in Eqs.(3.12) and (3.13). 
From Eqs.(3.21) and (3.23) ${\cal G}$ and $\cal F$ 
are expressed as 
\bea
&&{\cal G} 
\cong \pi {\hat{\ell}}^{-1}  \int _{1} ^\infty \frac{d v}{v^{2}} 
 \Delta ({t}_\ell v^{1/\nu}, {s}_\ell v^{\beta/\nu}).
\\
&& 
{\cal F} 
%\frac{\p}{\p {\hat\ell}}{\cal G} 
 \cong {\pi }{\hat{\ell}}^{-2}  \Delta 
({t}_{\ell}, {s}_{\ell}).
\ena 
We  introduce   two new scaling parameters,    
\bea 
&&t_\ell= 
\hat{t}\hat{\ell}^{1/\nu}=\tau (\ell/\xi_0)^{1/\nu},\nonumber\\
&& 
s_\ell=  \hat{s} \hat{\ell}^{\beta/\nu}=\psi_\infty/\psi_\ell,
\ena 
with  $\psi_\ell  =
  1.47 b_{\rm cx}(\xi_0/\ell)^{\beta/\nu}$. Here, $D$ in $t$ and $s$ 
is  replaced   by $\ell$ in $t_\ell$ and $s_\ell$.  
For small $\hat\ell$,    the products 
 $\hat{\ell}{\cal G} $ and  ${\hat{\ell}}^2 {\cal F} $ 
are determined only by   $t_\ell$ and 
$s_\ell$ from Eqs.(B4) and (B5). 

 We previously 
introduced another universal amplitude for near-critical  films, 
written as ${\cal A}(t,s)$, where   
the osmotic pressure is expressed 
as $\Pi=-k_BT {\cal A}/D^3$ \cite{OkamotoCasimir}. 
It is related to $\Delta(t,s)$ in three dimensions by  
\be 
 {\cal A}(t,s)= \bigg[2- \frac{\beta s}{\nu} \frac{\p }{\p s}- 
 \frac{t}{\nu} \frac{\p }{\p t}\bigg]\Delta(t,s) .   
\en 
We notice that differentiation of ${\cal F}$ 
in Eq.(B5) with respect to $\hat\ell$ at fixed 
$\hat t$ and $\hat s$ just yields  
\be 
\frac{\p}{\p \hat{\ell}} {\cal F} 
\cong -\pi {\hat{\ell}}^{-3}{\cal A}(t_\ell,s_\ell).
\en

We remark the following. 
(i) First, for small $t$ and $s$, 
 $\Delta(t,s)$ approaches its  critical-point value 
 $ \Delta_{\rm cri} =\Delta(0,0) \cong 
0.279 $ \cite{OkamotoCasimir}. 
Thus,  ${\cal G}$ 
and ${\cal F}$ grow  as in Eq.(3.26)  
for $\hat{\ell}\ll 1$. 
(ii) Second, for  $\tau=0$ (at $T=T_c$), 
we obtain 
\be 
{\hat{\ell}}^2 {\cal F}/\pi \cong 
 \Delta(0,s_\ell).
\en 
See Fig.4 of Ref.\cite{OkamotoCasimir} for 
 $\Delta(0,s)$. For ${\hat s} >0$,  
${\hat{\ell}}^2 {\cal F}/\pi$  decays from 
$\Delta_{\rm cri}$ to zero monotonously 
with increasing $\hat\ell$.  
For  $\hat{s}<0$, 
it takes a large maximum  about $3.73=13.4\Delta_{\rm cri}$ 
at $s_\ell=-0.90$ or at 
\be 
{\hat{\ell}}= 0.82 |\hat{s}|^{-\nu/\beta}= 6.14\xi/a,
\en 
where $\xi$ is defined by Eq.(2.9) (see the sentences 
below Eq.(3.13)).
%\xi /a= [1 / \sqrt{2 \delta(\delta+1)} ]
% \times | \hat{s} |^{-\nu / \beta} 
 This  explains the large maximum of ${\hat{\ell}}^2 {\cal F}$ 
for $(\hat{t}, \hat{s})=(0,-1)$ 
 in  Fig.3. (iii) 
Third, for $\psi_\infty=0$ and $\tau>0$ 
(on  the critical path), we obtain 
\be 
{\hat{\ell}}^2 {\cal F}/\pi\cong 
 \Delta(t_\ell,0). 
\en 
See Fig.8 of Ref.\cite{OkamotoCasimir} for 
 $\Delta(t,0)$.  For ${\hat t}>0$, ${\hat{\ell}}^2 {\cal F}/\pi$  starts  
from $\Delta_{\rm cri}$, takes  a mild maximum about 
$0.544=1.95\Delta_{\rm cri}$ at $t_\ell =2.30$ 
or at  
\be 
\hat{\ell} \cong 1.64\hat{t}^{-\nu}= 1.64\xi/a ,
\en   
and goes  to zero for larger  $\hat\ell$. This yields 
the mild minimum of  ${\hat{\ell}}^2 {\cal F}$ 
for $(\hat{t}, \hat{s})=(5,0)$ 
 in  Fig.3. 
%\hat{\ell}^2{\cal F}=1.71'Æ'È'Á'Ä'¢'Ü'·B
%'¿'È'Ý'ÉDerjaguin‹ßŽ—'Å'Í‹É'å'Í\hat{\ell}=0.613'Å‹É'å'É'È'èA'»'Ì'l'Í1.63'Å'·

Finally, we  discuss how  $\cal F$ and $\cal G$ 
behave  away from the criticality 
or for $|\hat{t}_\ell|\gg 1$ 
or   $|\hat{s}_\ell|\gg 1$. 
Our previous work \cite{OkamotoCasimir} 
indicates that   
if $|t| \gg 1$ or $|s|\gg 1$, ${\cal A}(t,s)$ decays as  
\be 
{\cal A}(t,s)\sim (D/\xi)^3 \exp(-D/\xi), 
\en 
where $\xi$ is defined by Eq.(2.9). 
% Note that $D/\xi$ is a function of 
% $t$ and $s$ from Eq.(B3).   
Let  the midplane   value of $\psi$ at $z=D/2$ be 
denoted by $\psi_m$ for a film. Then,   Eq.(B13) 
follows for $\psi_m \cong \psi_\infty$, where we hve 
  $-\ln(\psi_m/\psi_\infty-1) 
\sim D/2\xi$ and   ${\cal A}\cong 
D^3f''(\psi_\infty) (\psi_m-\psi_\infty)^2/k_BT_c$ \cite{OkamotoCasimir}. 
We now  need to replace $(t,s)$  by $(t_\ell,s_\ell) $ 
in Eq.(B13). To this end, we consider the combinations,  
\be 
q\equiv |\hat{t}|^\beta/\hat{s}= 
|{t}|^\beta/{s}=|{t}_\ell|^\beta/{s}_\ell= 
1.47b_{\rm cx}|\tau|^\beta/\psi_\infty,
\en 
 which 
do not depend on  $a$, $D$, and $\ell$.
We may assume the scaling relation 
 $\xi= \xi_0 |\tau|^{-\nu} M(q)$ 
in the critical region (for small 
$\tau$ and $\psi_\infty$), 
where $M(q)$ is a scaling function of $q$. 
Then, $D/\xi=  |t|^{\nu} M(q)^{-1}$ in Eq.(B13).
Replacement $(t,s)\to (t_\ell,s_\ell) $ 
simply  yields $D/\xi \to \ell/\xi$, leading to  
\be  
{\cal A}(t_\ell,s_\ell)\sim (\ell/\xi)^3 
\exp(-\ell/\xi).
\en   
Substitution of this relation 
into  Eqs.(B8) and use of Eq.(3.23) give   ${\cal F}$ 
and $\cal G$ in Eq.(3.27).


\begin{references}
\bibitem{Casimir} 
The Casimir interaction  was 
originally found to be induced 
by the ground-state fluctuations 
of the electromagnetic field between two mirrors.
In near-critical fluids, the solvent-mediated 
interaction   
is caused  by   preferential adsorption, 
where the composition heterogeneity is 
stationary for  fixed colloid positions.
From our viewpoint, it
is  misleading to call it the Casimir interaction,
Thus we  call it  the adsorption-induced 
 interaction \cite{Beysens}.
\bibitem{Casimir-comment} The coefficient $C_1$ 
in Ref.\cite{OkamotoCasimir}
 is set equal to 1 in the present paper. Then  
$C$ is dimensionless.
\bibitem{mid} For $\tau=0$, $\psi$ and 
$\omega_{\rm loc}$  are  calculated  as  $\psi_m  =  0.7107 \psi_D$ and 
 $\omega_m= 0.205 k_BT_c D^{-3}$, respectively, at $z=D/2$  
in the film geometry \cite{OkamotoCasimir}. 
See  Eq.(B3) for  $\psi_D $.
\bibitem{Gamb1} 
The authors of Ref.\cite{Gamb} obtained 
${\cal G}= \pi Aa\xi^{-1}\exp(-\ell/\xi)$ 
with $A=1.2-1.4$ for $\ell>\xi$ on the critical 
path ($\tau>0$ and $\psi_\infty=0$). 
In our scheme,   the curve of $\ln({\cal G})$ for 
$(\hat{t}, \hat{s})=(5,0)$ in Fig.6 gives $A\cong 0.9$.
\end{references}
\end{document}